\documentclass[aps,twocolumn,prd,superscriptaddress]{revtex4-2}

\usepackage[utf8]{inputenc}

\usepackage{mathtools}
\usepackage{amsfonts}
\usepackage{mathrsfs}
\usepackage{bbm}
\usepackage{physics}
\usepackage{slashed}
\usepackage{tensor}

\usepackage{graphicx}
\usepackage{color, float}
\usepackage{array}
\usepackage[abs]{overpic}

\usepackage{placeins}

\usepackage{makecell}
\usepackage{subcaption}

\usepackage{xspace}
\usepackage{siunitx}
\usepackage{xfrac}
\usepackage{hyperref}

\usepackage[nameinlink]{cleveref}

\makeatletter
\AddToHook{cmd/appendix/before}{\def\cref@section@alias{appendix}}
\AddToHook{cmd/appendix/before}{\def\cref@subsection@alias{appendix}}
\AddToHook{cmd/appendix/before}{\def\cref@subsubsection@alias{appendix}}
\AddToHook{cmd/appendix/before}{\def\cref@paragraph@alias{appendix}}
\makeatother
\crefname{appendix}{Appendix}{Appendices}
\Crefname{appendix}{Appendix}{Appendices}

\usepackage{bookmark}

\usepackage{xifthen}
\usepackage{xcolor}
\hypersetup{
	colorlinks,
	linkcolor={red!75!black},
	citecolor={blue!75!black},
	urlcolor={blue!75!black}
}

\usepackage{booktabs}
\usepackage{multirow}

\newcolumntype{C}{>{$}c<{$}}
\AtBeginDocument{
	\heavyrulewidth=.08em
	\lightrulewidth=.05em
	\cmidrulewidth=.03em
	\belowrulesep=.65ex
	\belowbottomsep=0pt
	\aboverulesep=.4ex
	\abovetopsep=0pt
	\cmidrulesep=\doublerulesep
	\cmidrulekern=.5em
	\defaultaddspace=.5em
}

\graphicspath{{./figures}}

\newcommand{\gettitle}{Solving fermionic sign problems with physics-informed kernels}

\newcommand{\getBochumAffiliation}{\affiliation{Theoretische Physik III, Ruhr-Universit{\"a}t Bochum, Universit{\"a}tsstraße 150, 44801 Bochum, Germany}}
\newcommand{\getHeidelbergAffiliation}{\affiliation{Institut f{\"u}r Theoretische Physik, Universit{\"a}t Heidelberg, Philosophenweg 16, 69120 Heidelberg, Germany}}
\newcommand{\getEMMIAffiliation}{\affiliation{ExtreMe Matter Institute EMMI, GSI, Planckstr.~1, 64291 Darmstadt, Germany}}
\newcommand{\getTokyoAffiliation}{\affiliation{Department of Physics, The University of Tokyo, 7-3-1 Hongo, Bunkyo-ku, Tokyo 113-0033, Japan}}

\hypersetup{
	pdftitle={\gettitle},
	pdfauthor={Fukushima, Ihssen, Kapust, Pawlowski},
	pdfkeywords=
	{physics-informed kernels} {PIKfan} {fermionic sign problem} ,
	bookmarksopen=true,
	bookmarksopenlevel=2,
	bookmarksnumbered=true
}

\begin{document}
	
\title{\gettitle}
    \author{Kenji Fukushima}
    \getTokyoAffiliation
	\author{Friederike Ihssen}
	\getBochumAffiliation
	\getHeidelbergAffiliation
	\author{Renzo Kapust}
	\getHeidelbergAffiliation
	\author{Jan M. Pawlowski}
	\getHeidelbergAffiliation
	\getEMMIAffiliation
\begin{abstract}
  We extend the generative sampling approach with physics-informed kernels (PIKs) \cite{Ihssen:2025ybn,Ihssen:2026njd} to a general class of 
  physics systems with complex distributions and a structure also encountered in fermionic systems. The resolution of sign problems of generic weighted sums of complex Gau\ss ian distributions with the PIK approach is central to this application. 
  In a first step, PIKs map a real normal distribution to such a weighted sum, and this procedure is called the PIKfan. The PIKfan has the  sampling cost of a real normal distribution. In a second step, this weighted sum is evolved with a second PIK into the distribution of the theory at hand. In combination, we reduce the sampling task for a general class of  
  complex systems to a Gau\ss ian one, and the main computational task is that of computing the PIK. We illustrate the approach with the one-site Hubbard model at arbitrary chemical potential and with $\phi^4$-deformed extensions of it.
\end{abstract}

\maketitle

\section{Introduction}
\label{sec:Intro}

Lattice field theory provides access to many interesting strongly correlated physics systems, and its numerical formulation is amenable to efficient sampling architectures. Still, even qualitative access eludes us in systems with sign problems, ranging from statistical and condensed-matter systems, mostly fermionic ones, to Quantum Chromodynamics (QCD) and Beyond Standard Model (BSM) physics. 
Moreover, real-time physics requires computing high-dimensional integrals with respect to a distribution that oscillates with a pure phase.

The sign problem may be NP-hard, see e.g.~\cite{Troyer:2004ge,Aarts:2026uiu}, and then even the most efficient sampling algorithm runs out of steam eventually if one tries to overcome the problem by brute force. Still, efficiency goes a long way as in most cases it suffices to delay the onset of the sign problem or its volume scaling in the continuum limit. Apart from efficiency, one can reparametrise the sampling task such that the sign problem is absent. Both aspects have been tried in the past. Prominent examples of reparametrisation approaches to the sampling tasks are sampling with complex Langevin equations (CLEs), see e.g.~\cite{Aarts:2017vrv,Boguslavski:2024yto,Fukushima:2024oij,Aarts:2025lpi,Alvestad:2025iep,Mandl:2026ngb,Boguslavski:2026rpx}, the Lefschetz thimble approaches, see e.g.~\cite{Cristoforetti:2012su,Fujii:2013sra,Tanizaki:2015rda,Alexandru:2015sua,Bluecher:2018sgj,Fukuma:2019wbv,Alexandru:2020wrj,DiRenzo:2020cgp,Ulybyshev:2024kdr,Fukuma:2025uzg}, optimised contour deformations, see e.g.~\cite{Mori:2017pne,Lawrence:2022dba,Lawrence:2023sfc,Gantgen:2023byf,Schuh:2026dvp}, and dual variables, see e.g.~\cite{Chandrasekharan:2013rpa,Gattringer:2014nxa,Gattringer:2016kco}. In all these approaches, efficient and accurate sampling is also of central importance. A prominent example of the direct combination of reparametrisation and  brute force is sampling with the density of states, see e.g.~\cite{Gattringer:2016kco,Francesconi:2019nph,Gattringer:2019khb,Pawlowski:2022rdn,Singh:2026zls}. Thus, practical solutions to the sign problem will most likely combine both aspects. 

In \cite{Ihssen:2025ybn, Ihssen:2026njd} a generative architecture for lattice field theories, sampling with physics-informed kernels (PIKs), was put forward, which naturally encompasses both aspects in a unified picture. It is based on a Fokker--Planck-type evolution or flow equation for the distribution $p(\varphi)$ of the theory under investigation. This flow equation, the Wegner equation \cite{Wegner_1974}, accommodates general reparametrisations and deformations of the distribution. The PIK is the differential map that encodes both the flow of the distribution and that of the sampling manifold. This formulation allows one to map the sampling task in a given theory to that in a simpler one, in most cases a real normal distribution. Moreover, 
it has been shown in \cite{Ihssen:2026njd} that Wegner flows have the following property, which is crucial for sampling theories with sign problems: they are \textit{weight-preserving} for general PIKs. This means that real sampling is always mapped to real sampling (no sign problem) and the probability weight of all intervals is preserved (no reweighting problem). This entails in particular that if one can construct a PIK that connects the sampling with a simple real positive distribution to that of a complex one, 
one can map the sampling of a sign problem to sampling a real simple distribution, needing no further structure. It is this property which sets sampling with PIKs qualitatively apart from other approaches to the sign problem such as the methods based on the CLEs, Lefschetz thimbles, the density of states, optimal manifolds, and normalising flows.

The above immediately raises the question whether one can always construct a \textit{real} PIK from some \textit{real} distribution to a given complex one of a theory with a sign problem. In \cite{Ihssen:2026njd}, a real PIK has been constructed for the simple real-time harmonic oscillator, hence solving the sampling task there within the PIK approach. 
The requirement for constructing a real PIK is a path of distributions whose zeros do not reach the evolving sampling manifold. 
This is the case for real-time scalar theories, whose distributions $e^{iS}$ have no zeros, and thus they are very likely solvable within the PIK approach. By now the proof of principle with the harmonic oscillator has been extended to interacting scalar systems~\cite{PIK2026}.

In the present work we generalise the complex PIK approach to theories that do not allow for a direct application of real sampling. We show that these theories can also be sampled from a real positive distribution, but the map to the physics distribution is done with a weighted sum of PIKs with positive relative weights. As shown in \cite{Ihssen:2026njd}, the weight-preserving property of each PIK maps real positive sampling to real positive sampling without reweighting, and hence without an overlap problem. With positive relative weights, this property carries over to the weighted sum, which resolves the sign problem. This approach is generally applicable and completes the conceptual setup of the PIK approach to sampling from complex distributions. We illustrate important properties of the setup using the example of a one-site Hubbard model. 
In the presently used formulation, the directly sampled model is plagued with a sign problem of varying severity at different chemical potentials, but admits an analytic solution. 
We show that the model can be sampled with the PIK approach in all regimes at the cost of Gau\ss ian sampling.

We close this Section with a bird's-eye view on the structure of the paper. In \Cref{sec:SolvingPIKs} we briefly recapitulate the PIK approach and its weight-preserving property, and show how it solves sign problems whenever a real PIK exists. We also show with a simple example how zeros of the distribution, as they typically occur in fermionic theories, obstruct the construction of such a real PIK. In \Cref{sec:PIKfan} we overcome this obstruction with the weighted sum of PIKs mentioned above, which we call the \textit{PIKfan}, and discuss how it can be set up constructively within the computation of the PIK itself. In the same Section, we also introduce the notion of equivalence classes $\mathcal{E}(p, \mathcal{M})$, namely classes of sampling pairs that can be mapped into each other without facing obstructions from irregular maps. In \Cref{sec:OneSiteHubbard} we apply the PIKfan to the one-site Hubbard model, also in the regime where a single real PIK fails, as well as to $\phi^4$-deformed extensions of the same. We conclude in \Cref{sec:Conclusions}, and the Appendices provide further details on the examples and on the role of the zeros.

%%%%%%%%%%%%%%%%%%%%%%%%%%%%%%%%%%%%%%
\section{Solving sign problems with PIKs}
\label{sec:SolvingPIKs}

We consider the problem of sampling from a normalised target (physics) distribution $p(\varphi)\in\mathbb{C}$ on a lattice $\mathcal{D}$ in $d$ spacetime dimensions with $\varphi\in\mathcal{M}$ on a sampling manifold ${\cal M}$. Formulated for expectation values of general operators ${\cal O}$ this is the sampling task 
\begin{align} 
	\langle  {\cal O}\rangle_{(p,{\cal M})}:=\int\limits_{{\cal M}} d\mu(\varphi) \,p(\varphi)\, {\cal O}(\varphi)\,, \qquad \langle  1 \rangle_{(p,{\cal M})}=1\,. 
\label{eq:SamplingProblem}
\end{align}
The pair $(p(\varphi)\,,{\cal M})$ is called the \textit{target sampling pair}. 
The PIK approach \cite{Ihssen:2025ybn, Ihssen:2026njd} reformulates the sampling task \labelcref{eq:SamplingProblem} with the target sampling pair as the real sampling task with the sampling pair $(p_0(\phi)\,,{\cal M}_0)$. Here, $p_0(\phi)$ is a 
\textit{chosen} simple normalised trial distribution and $\phi$ lives on a \textit{chosen} sampling manifold ${\cal M}_0$. The sampling pair is chosen such that it 
allows for sampling from a (real) probability distribution.  

In \Cref{sec:PIK-Architecture} we briefly recapitulate the main properties of the approach and in \Cref{sec:RealPIKfolds} we discuss how sign problems are solved within the PIK approach. Finally, in \Cref{sec:RippingExample} we discuss the limits of the direct application, which leads us to its generalisation in \Cref{sec:PIKfan}.

%%%%%%%%%%%%%%%%%%%%%%%%
\subsection{The PIK architecture} 
\label{sec:PIK-Architecture}

Within the PIK architecture, sampling from a target sampling pair $(p(\varphi)\,,{\cal M})$ is reformulated as a combination of three tasks: \\[-1ex] 

First, the sampling is done with a typically \textit{real} sampling pair $(p_0\,,\,{\cal M}_0)$ with a simple distribution $p_0(\phi)$, e.g.~with a normal distribution, drawing samples from the \textit{real} sampling manifold ${\cal M}_0$. \\[-2ex]
 
The second and main task is the (numerical) computation of the push-forward of the samples of the initial sampling pair $(p_0,{\cal M}_0)$ to that of the sampling pair $(p_1,{\cal M}_1)$ with the target distribution $p_1=p$,
\begin{align}
(p_0\,,\, {\cal M}_0)\longrightarrow  (p_t\,,\, {\cal M}_t)\longrightarrow  (p_1\,,\, {\cal M}_1)\,, 
	\label{eq:RealPIK}
\end{align} 
with the intermediate sampling pairs $(p_t\,,\,{\cal M}_t)$ with normalised distributions $p_t$ and $t\in [0,1]$.  
The PIK $\dot\phi_t$ is the differential map or push-forward and is computed with the Wegner equation \cite{Wegner_1974}. For that purpose it is convenient to parametrise the normalised distributions $p_t$ with an action $S_t$,
\begin{align}
	p_t = e^{-S_t}\,,
\end{align}
where $S_t$ is nothing but the negative log-likelihood of the distribution $p_t$.  We note that in addition to the field-dependent terms, $S_t$ also contains the log of the normalisation.
Then, the Wegner equation \cite{Wegner_1974} reads
\begin{align}
	\frac{\partial p_t}{\partial t}=-\frac{\partial}{\partial \phi} \left[ \dot\phi_t\,p_t \right] \  \longleftrightarrow \  
	\left(\frac{\partial}{\partial \phi} - \frac{\partial S_t}{\partial \phi}\right) \,\dot \phi_t =\frac{\partial S_t}{\partial t} \,. 
	\label{eq:PIKflow}
\end{align}
\Cref{eq:PIKflow} is simply a continuity equation for the distribution. For real distributions, the PIK $\dot\phi_t$ is the velocity field of the flow and $\dot\phi_t \,p_t$ is the probability current.\\[-2ex]

This leaves us with the third and final task: it has to be shown that sampling with the  sampling manifold ${\cal M}_1$ is equivalent to sampling with the sampling manifold ${\cal M}$ leading to \labelcref{eq:SamplingProblem}. This can be done by Cauchy's theorem, or more generally, it requires monitoring the non-analyticities of the integrand between the two manifolds. \\[-2ex]

\begin{figure}[t]
	\centering
	\includegraphics[width=1.\linewidth]{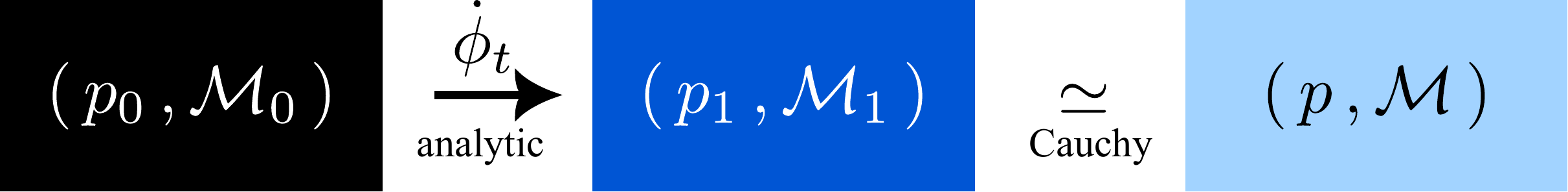}
	\caption{Illustration of the three key steps (1,2,3) in the PIK approach to computing observables for a complex distribution $p$ on the manifold $\mathcal{M}$: \hspace*{\fill} \vspace{.15cm}\newline 
		(1) Sampling with the sampling pair $(p_0, \mathcal{M}_0)$. \hspace*{\fill}\vspace{.1cm} \newline
		(2) Next, the pair is transported using PIKs $\dot\phi_t$ to $(p_1, \mathcal{M}_1)$, with $p_1 = p$, see \labelcref{eq:RealPIK}. The integrated PIKs $\phi_t(\phi)$ have to be analytic in a neighbourhood of the manifolds ${\cal M}_t$.\hspace*{\fill}\vspace{.1cm}\newline 
		(3) When Cauchy's theorem applies, the integral over $\mathcal{M}_1$ equals that over the target manifold  $\mathcal{M}$, see \labelcref{eq:CauchyM1M}. 
			\hspace*{\fill}}
	\label{fig:concept_single_pik_chain}
\end{figure}

The workflow for computing expectation values of the sampling pair $(p, \mathcal{M})$ via sampling from $(p_0, \mathcal{M}_0)$ is illustrated in \Cref{fig:concept_single_pik_chain}. \\[-1ex]

Following this brief outline, we provide further details on the PIK approach. To begin with, expectation values of operators ${\cal O}$ with the sampling pair $(p_t,{\cal M}_t)$ are given by 
\begin{align} 
	\langle  {\cal O}\rangle_{(p_t,{\cal M}_t)}:=\int\limits_{{\cal M}_t} d\mu_t(\phi) \,p_t(\phi)\, {\cal O}(\phi)\,, 
	\label{eq:ExpOt}
\end{align} 
with the normalised distributions $p_t$, 
\begin{align} 
	\int\limits_{{\cal M}_t} d\mu_t(\phi)\,p_t(\phi)=1\,. 
	\label{eq:Normalisation} 
\end{align} 
Here, $d\mu_t(\phi)$ is the tangential measure of the manifold ${\cal M}_t$ and $t \in [0,1]$. 

Importantly, the PIK $\dot\phi_t$ with \labelcref{eq:PIKflow} evolves both the distribution and the manifold: each configuration $\phi$ on the manifold ${\cal M}_t$ at flow time $t$ is pushed forward with the PIK, 
\begin{align} 
 \hspace{-.15cm} d \phi_t(\phi) = \dot\phi_t(\phi) d t\,,\quad \!{\cal M}_{t+dt} =\left\{\phi+ d\phi_t(\phi)
  | \phi\in  {\cal M}_t\right\}\!.
\label{eq:Mt+dt} 
\end{align} 
This entails that the measure $d\mu_t$ has the flow 
\begin{align} 
	\frac{d}{dt} d\mu_t = \frac{\partial\dot\phi_t(\phi)}{\partial \phi} d\mu_t(\phi)\,.
\label{eq:FlowMeasure} 
\end{align} 
While in practical numerical computations only the infinitesimal push-forward of configurations is needed, the integrated form also merits a discussion. The integrated form of \labelcref{eq:Mt+dt}  reads 
\begin{align} 
	{\cal M}_t =\left\{\phi_t(\phi)\,|\,\phi \in{\cal M}_0\right\}\,, 
	\label{eq:Mt} 
\end{align} 
with the integrated PIK, 
\begin{align}
	\phi_t(\phi) = \phi + \int_0^{t} ds \, \dot \phi_s (\phi_s)\,, \qquad \varphi(\phi) = \phi_1(\phi)\,.
	\label{eq:phiMap}
\end{align}
\Cref{eq:FlowMeasure} and the Wegner equation \labelcref{eq:PIKflow} encode the pivotal weight-preservation property of the PIK architecture: the density $d\mu_t \,p_t(\phi_t)$ at a given configuration $\phi_t$ for the sampling pairs $(p_t,{\cal M}_t)$ is $t$-independent,
\begin{align}\nonumber 
\frac{d}{d t} \left[ d\mu_t \,p_t(\phi_t) \right]  =&\, \frac{\partial\dot\phi_t(\phi)}{\partial \phi} d\mu_t(\phi)\, p_t(\phi_t) 
- d\mu_t\,\frac{\partial}{\partial \phi} \left[ \dot\phi_t\,p_t \right]\\[1ex]
 &\,+ d\mu_t \,\dot \phi_t \frac{\partial p_t}{\partial \phi}[\phi_t]=0\,,
\label{eq:WP}
\end{align}
where the three terms on the right-hand side correspond to the $t$-derivatives of $d \mu_t$, $p_t$ and $\phi_t$, respectively.

If the map 	\labelcref{eq:phiMap} is analytic, the expectation value of general operators ${\cal O}$ with the sampling pair $(p,{\cal M}_1)$ can be computed with the weight-preserving property \labelcref{eq:WP} of the PIK with the initial sampling pair $(p_0,{\cal M}_0)$, 
\begin{align} 
	\int\limits_{{\cal M}_0} d\mu_0 \,p_0(\phi) \,{\cal O}(\varphi(\phi)) =  \int\limits_{{\cal M}_1} d\mu_1 \,p(\phi) \,{\cal O}(\phi)  \,,
	\label{eq:PIKsampling}
\end{align} 
with $\varphi(\phi)$ in \labelcref{eq:phiMap}. Note that \labelcref{eq:PIKsampling} holds for all $t$ if $\phi_t(\phi)$ is used on the left-hand side. This is pictured by step (2) in \Cref{fig:concept_single_pik_chain}.

Importantly, the requirement that the map from \labelcref{eq:phiMap} must be analytic can be directly checked in the PIK approach. This is a feature of the constructive nature of PIKs and is discussed in detail in \Cref{sec:NoRip,sec:zeros_and_transport}. In particular, by tracking the modulus of the integrated change in the action
\begin{align}
	\Lambda_t(\phi_0) := \left| \int_0^t ds\, \frac{\partial\dot\phi_{s,j}(\phi_s)}{\partial \phi_{j}} \right|\,,
	\label{eq:regularity_indicator}
\end{align}
where $j$ labels the lattice points and is summed over, and $\phi_s$ is the solution of \labelcref{eq:phiMap} with initial condition $\phi_0$, one obtains a scalar indicator of the map's regularity that is connected to the evolution of the manifold, see e.g.~\labelcref{eq:FlowMeasure}. Moreover, as discussed in \Cref{sec:zeros_and_transport}, the analytic access to the Wegner equation relates irregular maps to features of the distribution $p_t$. For the remainder of this work, we show how this analytic access can be used to constructively set up sampling methods that ensure regular maps.

Given the analytic map $\varphi(\phi)$, \labelcref{eq:PIKsampling} maps the original sampling task with the sampling pair $(p,{\cal M})$ to sampling with the initial sampling pair $(p_0,{\cal M}_0)$, if the sampling manifold ${\cal M}_1$ can be deformed smoothly into the target manifold ${\cal M}$ without changing the integral. This step requires Cauchy's theorem to hold and hence we should not sweep over singularities of the integrand $p_1\,\mathcal{O}$ within the deformation. For the entire distributions considered in this work, these are singularities of the operators $\mathcal{O}$; for the PIKfan, singularities of the rib distributions $\rho_{i,1}$ also enter, see \Cref{sec:PropertiesOfPIKfans}. This is illustrated by step (3) in \Cref{fig:concept_single_pik_chain}. We also note in passing that more generally we can also collect the contributions of these singularities. While this is an interesting option, we shall not consider this further in this work.  

Then, computing the observable with respect to $p_1$ on the manifold ${\cal M}_1$ is the same as computing it with respect to $p$ on ${\cal M}$, 
\begin{align} 
	\int\limits_{{\cal M}_1} d\mu_1 \,p_1(\phi) \,{\cal O}(\phi)= \int\limits_{{\cal M}} d\mu \,p(\varphi) \,{\cal O}(\varphi)  \,.
	\label{eq:CauchyM1M}
\end{align} 
In short, using \labelcref{eq:PIKsampling} and \labelcref{eq:CauchyM1M}, sampling with the target pair $(p,{\cal M})$ can be done with the initial sampling pair $(p_0,{\cal M}_0)$. We shall see below that not all sampling pairs can be mapped into each other, see \Cref{sec:RippingExample}, which will lead us to the notion of equivalence classes, see \Cref{sec:EquivalentSampling}.

In any case, the setup above entails that with \labelcref{eq:PIKsampling} and \labelcref{eq:CauchyM1M} expectation values can be computed by sampling with the simple real distribution $p_0$ on ${\cal M}_0$,  
\begin{align}
	\langle \mathcal{O}\rangle_{(p, \mathcal{M})} \simeq \frac{1}{N_s} \sum_{n=1}^{N_s} \mathcal{O}\left(\varphi(\phi_n)\right) \,,
	\label{eq:SampleOtPIK}
\end{align}
where $\phi_n \sim p_0$ are drawn from ${\cal M}_0$. 

\Cref{eq:SampleOtPIK} requires the integrated map $\varphi(\phi)$ in \labelcref{eq:phiMap}. However, this map is constructed by integrating the infinitesimal map \labelcref{eq:Mt+dt} on the respective manifolds ${\cal M}_t$. The resulting manifolds are given by \labelcref{eq:Mt}. For a numerical implementation this entails the following: we draw a set of configurations ${\cal S}_0$ with the sampling pair $(p_0\,,\,{\cal M}_0)$. This is a discrete approximation of this manifold and the density of points is the probability density $d\mu_0\, p_0$. With \labelcref{eq:Mt+dt} the set ${\cal S}_0$ is evolved to the sets ${\cal S}_t$ and the density of points is that of the probability density $d\mu_t\, p_t(\phi_t) =d\mu_0\, p_0(\phi_0)$ with $\phi_t=\phi_t(\phi_0)$. Sampling with the sampling pairs $(p_t,{\cal M}_t)$ is simply done by summing over the configurations $\phi_t\in {\cal S}_t$. Hence, sampling with a given sampling pair and with \labelcref{eq:phiMap} and \labelcref{eq:SampleOtPIK} is simply given by 
\begin{align}
	\langle \mathcal{O}\rangle_{(p_t,{\cal M}_t)} \simeq \frac{1}{N_s} \,\sum_{\phi_{n}\in{\cal S}_{t}}  \mathcal{O}\left(\phi_{n}\right) \,. 
	\label{eq:SampleOtpMt}
\end{align}
For an exact PIK, sampling is thus done with the real and positive distribution $p_0$ and samples are pushed forward, which circumvents the direct sign problem. The complex map $\varphi(\phi)$ can still in principle enhance the variance of specifically chosen observables. For typical observables such as moments of the distribution, however, $\mathcal{O}(\,\varphi(\phi)\,)$ is merely a product of mapped fields, and no exponential growth of the variance with the volume is expected.

This concludes our discussion of the PIK architecture.

%%%%%%%%%%%%%%%%%%%%%%%%%%%%%%%%%%%%%%
\subsection{Solving sign problems with real PIKfolds}
\label{sec:RealPIKfolds}

We now briefly describe how the PIK architecture can be used to solve sign problems. For this purpose it is convenient to introduce the notion of a \textit{PIKfold} denoted by ${\cal P}$, which is the set of all sampling pairs $(p_t, {\cal M}_t)$ for $t\in[0,1]$, 
\begin{align} 
	{\cal P} := \left\{ (p_t\,,\,{\cal M}_t) \,| t\in[0,1]\right\} \,,
\label{eq:PIKfold} 
\end{align} 
and the name derives from the set of mani\textit{folds} in ${\cal P}$. For complex distributions $p_t$ the  manifolds ${\cal M}_t$ are embedded in the complexification ${\cal C}$ of ${\cal M}_0$, which is typically chosen to be the sampling manifold of the target distribution,
\begin{align} 
	{\cal M}_0={\cal M}\,.
\label{eq:StandardChoiceM0}
\end{align} 
This choice is particularly useful if ${\cal M}$ is a simple manifold such as $\mathbb{R}^\mathcal{D}$, which is typically the case in physical sampling problems.
In view of the sampling with the sets ${\cal S}_t$ as discussed above \labelcref{eq:SampleOtpMt}, we find that the set ${\cal S}$ of all ${\cal S}_t$ is a discrete approximation of the PIKfold ${\cal P}$ but also carries the probability density $d\mu_t\,p_t$. In short, it is the discrete approximation of the set of sampling pairs.

 Then, the solution of sign problems is centred on the weight-preserving property of the PIK. Let us start from an initial \textit{real} sampling pair $(p_0\,,\,{\cal M}_0)$,  
\begin{align} 
	d\mu_0 \,p_0(\phi) \in \mathbb{R}^+\,,\qquad \forall   \phi\in {\cal M}_0\,.
\label{eq:RealDensity}
\end{align}
Note that \textit{real} has to be read as \textit{real positive}. Now we assume that the solution $\dot\phi_t$ of the Wegner equation \labelcref{eq:PIKflow} is \textit{analytic} in a neighbourhood of the manifolds ${\cal M}_t\subset {\cal C}$. Then, the manifold ${\cal M}_t$ can be evolved further and Cauchy's theorem holds for this infinitesimal step. In its form for the distribution $p_t$, the Wegner equation \labelcref{eq:PIKflow} is read as a generalised Fokker--Planck equation. On these grounds, the embedding of other sampling approaches to the sign problem as specific PIKs has been discussed in \cite{Ihssen:2026njd}. Finally, the sampling with all sampling pairs $(p_t\,,\,{\cal M}_t)$ is done with a real positive probability density 
\begin{align} 
	d\mu_t \,p_t(\phi_t) = d\mu_0 p_0(\phi_0) \in \mathbb{R}^+\,,\qquad \phi_t=\phi_t(\phi_0)\,. 
\label{eq:RealProbabilityDensity} 
\end{align} 
This defines a \textit{real PIKfold} and as above it has to be read as \textit{real positive}: all sampling pairs have real positive and normalised densities \labelcref{eq:RealProbabilityDensity}, which are nothing but the initial simple probability density. In short, we have reduced sampling a complex distribution to sampling a simple probability distribution.  

We close this subsection with a few remarks relating the PIK approach to other methods tackling the sign problem. An important aspect of the PIK architecture is its generality. After all, it describes general transformations and reparametrisations of general (also complex-valued) distributions, enabling weight-preserving transport. Particular choices of the path $p_t$ or of the kernel $\dot\phi_t$ reproduce other known approaches. We outline these relations for some examples below; more details can be found in~\cite{Ihssen:2026njd}.

\textit{Complex Langevin:} For a particular choice of the kernel $\dot\phi_t$, the Wegner equation becomes equivalent to the Fokker--Planck equation used in complex Langevin approaches, see e.g.~\cite{Song:2020hus,Ihssen:2026njd}. For this choice of kernel, the deterministic flow of the PIK and the stochastic evolution of the fields pass through the same marginal distributions $p_t$. Hence, while the PIK approach can be seen as a non-equilibrium sampler, one can investigate the properties of complex Langevin methods with the analytic access of the PIKs.

\textit{Lefschetz thimbles:} Similarly, as discussed in~\cite{Ihssen:2026njd}, when choosing the kernel to follow the holomorphic gradient flow, $\dot\phi_{t,i}(\phi) = \overline{\partial_{\phi_i} S(\phi)}$, starting from the initial sampling manifold $\mathcal{M}$, the flow maps out the Lefschetz thimbles of the distribution. Sampling on multiple thimbles is generally expensive and suffers from a residual sign problem from different phases carried by different thimbles. As was shown in~\cite{Ihssen:2026njd}, by allowing different paths $p_t$ and kernels, the PIK approach enables one to sample efficiently on one contour where thimbles would require multiple.

\textit{Optimal manifolds:} Similarly, when optimising sampling manifolds in order to reduce the severity of the sign problem, both finding the manifold and sampling on it are generally expensive. In particular, one may find an optimal manifold but not know an efficient way to sample on it. Importantly, for PIKs finding the manifold and providing real and efficient sampling on it is done simultaneously and is provided by solving the Wegner equation, which is straightforwardly applicable also for complex-valued distributions.

Thus, as suggested by these examples, PIKs provide a general framework that can build on and improve existing sampling schemes while also describing new ones. This is particularly true for fermionic systems where the zeros of the distribution may lead to divergent drifts in complex Langevin approaches, see e.g.~\cite{Nishimura:2015pba,Aarts:2016mso,Aarts:2016bdr}, many contributing Lefschetz thimbles~\cite{Kanazawa:2014qma,Tanizaki:2015rda}, or non-analyticities in optimal manifold approaches. As we will show below, PIKs can use the analytic access to both the distribution $p_t$ and the infinitesimal field transformation $\dot\phi_t$ to track issues pertaining to the zeros of the distribution and build constructive approaches to avoid them.

%%%%%%%%%%%%%%%%%%%%%%%%%%%%%
\subsection{Ripping manifolds and Cauchy's theorem}
\label{sec:RippingExample}

There are interesting theories, such as many fermionic systems, that do not admit a real PIKfold, and the sampling procedure with PIKs has to be generalised. This happens when the sampling manifold hits the zero set ${\cal Z}_t$ of the distribution,
\begin{align}
	{\cal Z}_t\cap{\cal M}_t\neq\emptyset\,, \qquad {\cal Z}_t=\left\{\phi\in{\cal C}\,|\,p_t(\phi)=0\right\}\,.
\label{eq:ptZero}
\end{align}
In short, as $d\mu_t\,p_t=d\mu_0\,p_0$ stays finite by the weight-preserving property \labelcref{eq:WP} of the PIK, the measure $d\mu_t$ diverges at ${\cal Z}_t\cap{\cal M}_t$. The sampling manifold is stretched to infinity, and the derivative $\partial_\phi\dot\phi_t$ of the PIK diverges, see \labelcref{eq:FlowMeasure}. This leads to a \textit{ripping} of the manifold. 

These considerations, based on the weight-preserving property of the PIK, already show the general mechanism behind obstructions that measure-transport approaches may face for fermionic systems. They also indicate that PIKs can predict on their own whether their direct application requires generalisation. A more detailed discussion of the ripping for lattices with several sites, $|\mathcal{D}|\geq2$, is found in \Cref{sec:zeros_and_transport}.

In the following, we use the tractable zero-dimensional case to illustrate the ripping of the manifold in great detail. To that end, we first exemplify the ripping condition for general distributions $p_t$ in zero dimensions. Secondly, we provide an explicit example with the distribution found in \labelcref{eq:distribution_D_problem}.

To begin with, for the zero-dimensional case, the condition \labelcref{eq:ptZero} can be made explicit, as weight preservation fixes the PIKfold completely. In particular, ${\cal M}_t$ is a curve $\phi_t(\phi_0)$ with $\phi_0\in{\cal M}_0=\mathbb{R}$, and $d\mu_t\,p_t=d\mu_0\,p_0$ yields

\begin{align}
	f_t\bigl(\,\phi_t(\phi_0)\,\bigr)=f_0(\phi_0)\,,\quad f_t(\phi)=\int_{-\infty}^{\phi}d\phi'\,p_t(\phi')\,,
	\label{eq:CDFmap}
\end{align}
where $f_t$ is the cumulative distribution function (CDF) of $p_t$. Hence, ${\cal M}_t$ is the curve on which $f_t$ is real and increases monotonically from 0 to 1. In particular, the map $\phi_t$ can be computed from the CDF via
\begin{align}
	\phi_t(\phi_0)=f_t^{-1}(\,f_0(\phi_0)\,)\,,
	\label{eq:InverseCDFmap}
\end{align}
and the tangent is given by
\begin{align}
	\frac{\partial}{\partial\phi_0} \, \phi_t(\phi_0) = \frac{p_0(\phi_0)}{p_t(\phi_t(\phi_0))}\,.
	\label{eq:tangent_phi_t}
\end{align}
As $p_0(\phi_0) > 0$ by construction, the sign of $\textrm{Re}\,p_t$ on ${\cal M}_t$ determines whether $\textrm{Re}\,\phi_t(\phi_0)$ increases or decreases with increasing $\phi_0$. In particular, for $\textrm{Re} \, p_t < 0$, the point on the manifold $\phi_t(\phi_0)$ runs backwards w.r.t.\ $\phi_0$, for $\textrm{Re}\, p_t = 0$ it is parallel to the imaginary axis, and for $\textrm{Re}\, p_t > 0$ it runs forwards.

However, neither running backwards nor moving parallel to the imaginary axis is an inherent obstruction to the estimation of expectation values. An obstruction only arises at the zeros of $p_t$, which are the critical points of $f_t$, see \labelcref{eq:CDFmap}. Close to a zero $z_t$ of order $m$,
\begin{align}
	f_t(\phi)-f_t(z_t)\propto(\phi -z_t)^{m+1}\,.
\end{align}
Thus, $f_t^{-1}$, and therefore $\phi_t$, has a branch point at the critical value $f_t(z_t)$. The samples, labelled by their real (initial) CDF value, cannot pass through this branch point: ${\cal M}_t$ only bends around $z_t$ as long as $f_t(z_t)\notin(0,1)$. The manifold rips at the first time $t_\textrm{rip}$ at which the critical value crosses the interval of labels,
\begin{align}
	\mathrm{Im}\,f_t(z_t)=0\,,\qquad 0<\mathrm{Re}\,f_t(z_t)<1\,,
	\label{eq:RipCriterion}
\end{align}
that is, when the zero lies on ${\cal M}_t$. Then, the zero cuts through the manifold: close to $z_t$, every value of $f_t$ has $m+1$ pre-images, and the samples cannot cross the zero. Following their pre-images continuously in $t$, samples with $f_0(\phi_0)<\mathrm{Re}\,f_t(z_t)$ and $f_0(\phi_0)>\mathrm{Re}\,f_t(z_t)$ end up on different branches of $f_t^{-1}$. Hence, ${\cal M}_t$ is torn open at $z_t$, and the samples no longer reproduce the expectation values of $p_t$, see \Cref{sec:NoRip}.

Moreover, \labelcref{eq:CDFmap} allows us to exclude a real PIKfold to $p$ from the distribution alone, without specifying a path $p_t$. At $t=1$, the samples have to lie on the curve ${\cal M}_1=\phi_1(\mathbb{R})$ with $\phi_1=f_1^{-1}\circ f_0$, on which the CDF $f_1$ of $p$ is real and increases from 0 to 1. Hence, a real PIKfold to $p$ can only exist if this curve connects $-\infty$ with $+\infty$ without passing through a zero of $p$.

Already a simple weighted sum of Gau\ss ians is enough to produce zeros close to the sampling manifold, and such sums occur in the Hubbard model and similar fermionic theories. In the one-site Hubbard model the zeros are double, $m=2$, see \Cref{sec:PolesAndPiks,sec:PIKfanCauchy}.

We illustrate the ripping with a minimal zero-dimensional example with a single variable, for which ${\cal Z}_t$ is a single point. It serves as guidance for \Cref{sec:PIKfan}, where we construct a PIK approach for systems that do not admit a real PIKfold. The example is given by the path
\begin{align}
		p_t(\phi) = g(\phi) \left( 1 +  i t\,\frac{ \phi}{a} \right) \,, \qquad	g(\phi) = \frac{1}{\sqrt{2\pi}} e^{-\frac{\phi^2}{2}} \,.
		\label{eq:distribution_D_problem}
\end{align}
A similar example was already used to illustrate issues of CLE approaches in the presence of zeros, see e.g.~\cite{Aarts:2017vrv}. Evidently, $p_t$ is a normalised complex distribution on $\mathcal{M} = \mathbb{R}$ with a zero at $z_t=i a /t$. This zero moves from an infinite imaginary value at $t=0$ to a finite imaginary value as $t$ is increased. Note that as $p_0(\phi) = g(\phi)$, the initial distribution does not have a zero. The zero appears only for $t>0$ in the complex plane and hits the corresponding PIKfold at some finite time $t_{\text{rip}}$. 

\begin{figure}[t]
	\centering
	\includegraphics[width=0.475\textwidth]{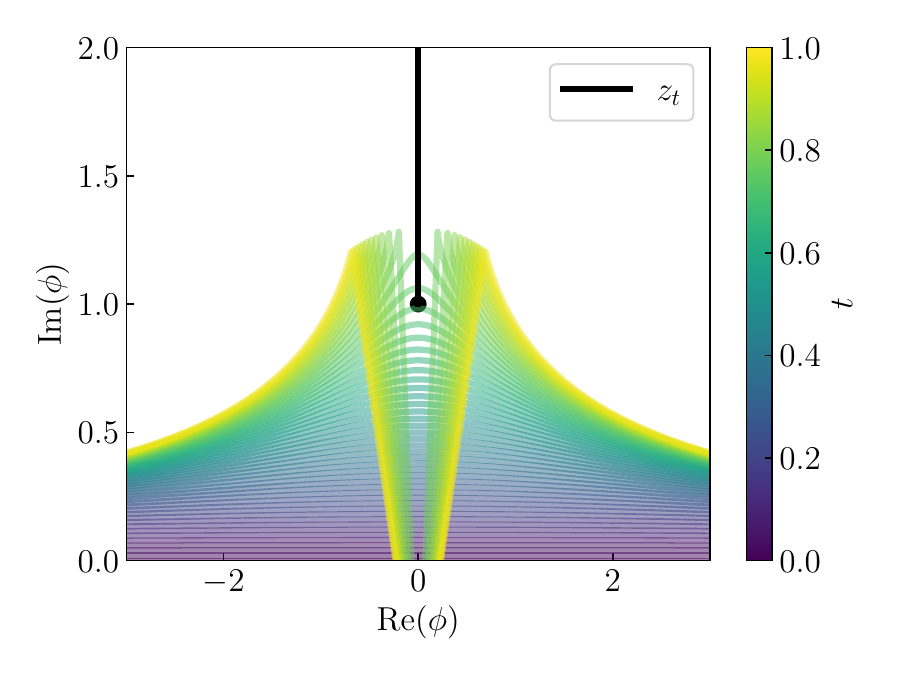}
	\caption{Sampling manifolds $\mathcal{M}_t$ computed from \labelcref{eq:phiMap}  and the path defined by \labelcref{eq:distribution_D_problem}.  The RG-time evolution is initialised at $\mathcal{M}_0 = \mathbb{R}$. The black line shows the movement of the zero $z_t$ of $p_t$ in the complex plane, for the choice $a=1$. The manifold rips open at $t_\textrm{rip}/a = 0.7651\ldots$, see \labelcref{eq:trip}.  
		\hspace*{\fill}}
	\label{fig:individual_zero_rip}
\end{figure}

In this simple model, the Wegner equation is solved analytically by
\begin{align}
	\dot \phi_t(\phi) = \frac{i}{a}\frac{1}{1 + i t\,\frac{\phi}{a}} \,,
	\label{eq:D_problem_kernel}
\end{align}
which has a pole at the zero $ z_t=i a /t$ of $p_t$. 
Nevertheless, we can use \labelcref{eq:phiMap} to compute the respective manifolds $\mathcal{M}_t$, starting at  $\mathcal{M}_0 = \mathbb{R}$. The evolving manifolds are shown in \Cref{fig:individual_zero_rip}. The black line indicates the movement of $z_t$ with varying $t$. When the evolving manifold $\mathcal{M}_t$ approaches the zero $z_t$, the manifold rips, signalling the failure of the real PIKfold.
Its CDF is $f_t(\phi)=G(\phi)-\frac{it}{a}\,g(\phi)$, where $G$ is the CDF of $g$. At the zero $z_t=ia/t$ we have $\mathrm{Re}\,f_t(z_t)=1/2$, and \labelcref{eq:RipCriterion} yields the time $t_{\rm rip}$ at which the manifold rips open as the positive real solution of
\begin{align}
  \frac{t_{\rm rip}}{a} = \sqrt{\frac{\pi}{2}} e^{-\frac{a^2}{2t_{\rm rip}^2}} \mathrm{Erfi}\left( \frac{a}{\sqrt{2}t_{\rm rip}} \right) \,,
  \label{eq:trip}
\end{align}
leading to $t_\textrm{rip}/a = 0.7651\ldots$. As $p_t'(z_t)=\frac{it}{a}\,g(z_t)$ with $g(z_t)>0$, the rays through $z_t$ point in the directions $\pm\pi/4$. Accordingly, $\mathrm{Re}\,p_t$ stays positive on ${\cal M}_t$ up to $t_\textrm{rip}$, and the manifold rips without ever becoming parallel to the imaginary axis.
Further details for this example are shown in \Cref{sec:NoRip}. This includes the sampling of the two-point function shown in \Cref{fig:E_problem_observable}, which also fails as the zero hits the manifold ${\cal M}_t$. 

%

%%%%%%%%%%%%%%%%%%%%%%%%%%%%%%%
\section{General complex PIK architectures: the PIKfan}
\label{sec:PIKfan}

The termination of the real PIKfold at a given $t_\textrm{rip}<1$ leaves us with the task of generalising our sampling procedure. To that end we first classify sign problems with the help of PIKs in \Cref{sec:EquivalentSampling} and use this classification to set up a sampling procedure  for general theories in \Cref{sec:RealSampling}.

%%%%%%%%%%%%%%%%%%%%%%%%%%%%%%%
\subsection{Equivalence classes of sampling pairs}
\label{sec:EquivalentSampling}

We start the classification task by illustrating it with a discussion of our simple example \labelcref{eq:distribution_D_problem} with the sampling pair $(p,\mathbb{R})$, where the target distribution is $p_1(\phi)=p(\phi)$ with 
\begin{align}
	p(\phi) = g(\phi) \left( 1 +  i \,\frac{ \phi}{a} \right) \,. 
	\label{eq:target}
\end{align}
Sampling with this sampling pair is equivalent to sampling with the sampling pairs $(p_t,{\cal M}_t)$ with $t\in(t_\textrm{rip},1]$, where the manifolds ${\cal M}_t$ are obtained by flowing backwards from $(p_1,{\cal M}_1)=(p,\mathbb{R})$. These are not the forward manifolds of \Cref{fig:individual_zero_rip}, which rip at $t_\textrm{rip}$: in $d=0$ they are the branch of the curve $\mathrm{Im}\,f_t=0$ on the other side of the zero $z_t$, see \labelcref{eq:CDFmap}.
Note that this does not define a real PIKfold as the original sampling problem is not real. 
We denote this equivalent sampling with 
\begin{align} 
(p_t,{\cal M}_t) \simeq (p,\mathbb{R})\,,\quad \textrm{for} \quad t\in(t_\textrm{rip},1]\,.
\label{eq:SimpleEquivalence}
\end{align} 
This equivalence has an important consequence: if we solve the sampling problem for one of the sampling pairs in this equivalence class, we have solved it for all since they are connected by a PIKfold. 

More generally, \textit{all} sampling pairs that can be obtained (in a singularity-free way) from the given sampling pair of a target distribution $p$ and the corresponding sampling manifold ${\cal M}$ either with a PIK $\dot \phi_t$ or by a deformation of the sampling manifold with a fixed $p$ using Cauchy's theorem define an equivalence class of sign problems: if we can solve the sign or sampling problem for a particularly simple sampling pair in this equivalence class, we have solved the sign or sampling problem for all sampling pairs within the same equivalence class. More specifically, we define the equivalence class by the following three equations: 
\begin{subequations} 
\label{eq:PIKCauchyEquivalenceClass}
\begin{align} 
	{\cal E}(p,{\cal M}_p)=\left\{ (q,{\cal M}_q)\,|\,   (q,{\cal M}_q)\simeq  (p,{\cal M}_p) \right\}\,,
	\label{eq:PIKCauchyEquiv}
\end{align} 
and the two sampling pairs are equivalent if there is a PIK $\dot \phi_t$ such that  
\begin{align} 
(p,{\cal M}_p) \stackrel{\dot\phi_t}{\longrightarrow}  (q,{\cal M}_1) \,,
\label{eq:PIKEquiv}
\end{align} 
and if we can invoke Cauchy's theorem: sampling with $q$ on the manifold ${\cal M}_1$ is the same as sampling with $q$ on the manifold ${\cal M}_q$: 
\begin{align} 
	\int\limits_{{\cal M}_1} d\mu_1 \,q(\phi) \,{\cal O}(\phi)=  \int\limits_{{\cal M}_q} d\mu \,q(\phi) \,{\cal O}(\phi)  \,.
	\label{eq:CauchyM1Mq}
\end{align} 
\end{subequations}
\Cref{eq:PIKCauchyEquivalenceClass} allows us to classify sign problems with the equivalence class of the physics distribution $p(\phi)$ and the physics sampling manifold ${\cal M}$. For invertible flows, the reverse flow, given by $-\dot\phi_{1-t}$, together with an application of Cauchy's theorem ensures that the equivalence relation is symmetric: $(p,{\cal M}_p)\simeq(q,{\cal M}_q)$ implies $(q,{\cal M}_q)\simeq(p,{\cal M}_p)$. The simplest equivalence class is given by sampling pairs $(p\,,\,{\cal M})$ that admit a real PIKfold and we call it ${\cal E}_\textrm{real}$. 

In the following we consider physical sampling pairs $(p\,,\,{\cal M})$ in the equivalence class of weighted sums of complex normalised distributions $\rho_i$ in the real equivalence class for all $i$,
\begin{align}
	( p,{\cal M})\in {\cal E}(p^{(w)},{\cal M})\,,
	\label{eq:Equivppw}
\end{align}
with 
\begin{subequations} 
\label{eq:WeightedSumDistribution}
\begin{align}
p^{(w)}(\phi)=	\sum_{i=1}^N w_i \, \rho_{i}(\phi) \,,
	\label{eq:WeightedSumDistribution1}
\end{align}
with 
\begin{align}
\sum_{i=1}^N w_i=1\,,\qquad \textrm{and}\qquad (\rho_i,{\cal M}) \in{\cal E}_\textrm{real}\,. 
	\label{eq:WeightedSumDistribution2}
\end{align}
\end{subequations} 
The $w_i\in \mathbb{R}^+$ are the relative positive weights of the corresponding \textit{complex} distributions $\rho_i$.
General principles for the construction of the $w_i$ and $\rho_i$ will be provided in \Cref{sec:FindingPIKfans}.
The respective PIK denoted by $\dot{\phi}_t^{(w)}$ is given by 
\begin{align} 
	(p^{(w)}\,,\,{\cal M}) \stackrel{\dot\phi_t^{(w)}}{\longrightarrow}	(p_1\,,\,{\cal M}_1)\,,
	\label{eq:PIKw} 
\end{align}
with $p_0=p^{(w)}$ and $p_1=p$. 
Note that typically the distributions $p^{(w)}(\phi)$ in \labelcref{eq:WeightedSumDistribution} have sign problems themselves if the final one has a sign problem. However, the sign problem of weighted sums can be solved with PIKs as demonstrated next in \Cref{sec:RealSampling}.

%%%%%%%%%%%%%%%%%%%%%%%%%%%%%%%
\subsection{The PIKfan: real sampling with weighted sums}
\label{sec:RealSampling}

\begin{figure}[t]
	\centering
	\includegraphics[width=.8\linewidth]{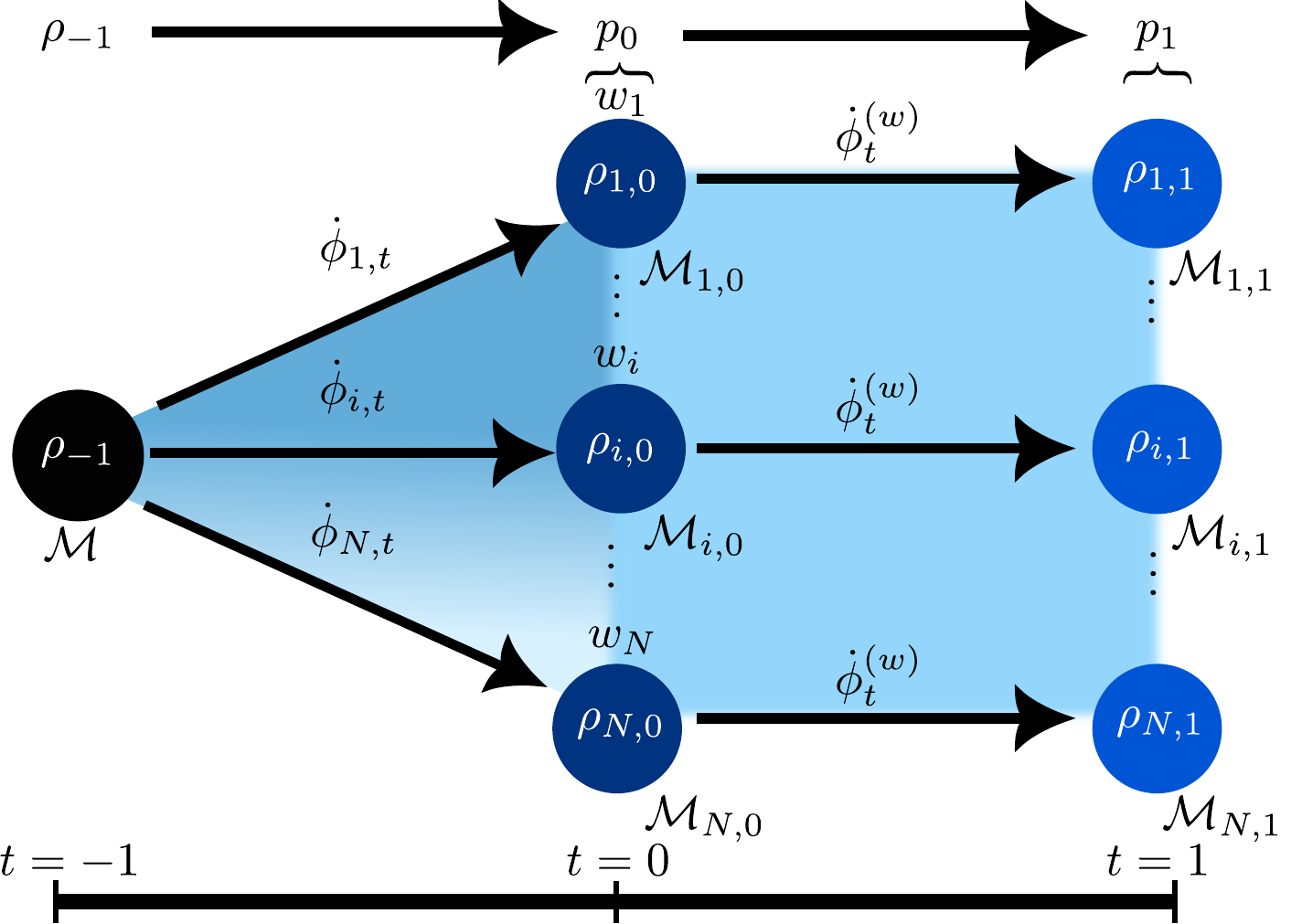}
	\caption{Illustration of the \textit{PIKfan}: A set of PIKs $\dot\phi_{i,t}$ maps the real normal sampling pair $(\rho_{-1}, \mathcal{M})$ to their respective (complex) sampling pairs $(\rho_i, \mathcal{M}_{i,0})$, which contribute with the weight $w_i$ to the distribution $p_0 = p^{(w)}$. A subsequent PIK $\dot\phi_t^{(w)}$ maps the sampling pairs at $t=0$ to $(\rho_{i,1}, \mathcal{M}_{i,1})$, which form the mixture that models the target distribution $p_1$. \hspace*{\fill}}
	\label{fig:ComplexRealPIKfan}
\end{figure}

By construction, all distributions $\rho_i$ are in the real equivalence class, $(\rho_i,\,{\cal M}) \in{\cal E}_\textrm{real}$. They admit real PIKfolds ${\cal P}_{i}$ where we choose $t\in[-1,0]$ with $\rho_{-1}=\rho_\text{real}$, where $\rho_\text{real}$ is some initial \textit{real normal} distribution, combined with a real sampling manifold ${\cal M}_{-1}={\cal M}$ of the physical theory with the sampling pair $(p,{\cal M})$,
\begin{align} 
	 (\rho_{-1}\,,\,{\cal M}) \stackrel{\dot\phi_{i,t}}{\longrightarrow } 	( \rho_{i,t}\,,\,{\cal M}_{i,t})\,,\quad \textrm{with}\quad \rho_{i,0}=\rho_i\,,
\label{eq:rho-1t}
\end{align} 
see \Cref{fig:ComplexRealPIKfan}. The choice ${\cal M}_{-1}={\cal M}$ is not required, but we want to keep the construction as simple as possible; see \labelcref{eq:StandardChoiceM0}. The definition of the real PIKfold and the condition of $(\rho_i,{\cal M})\in{\cal E}_\text{real}$ entail that we can deform the final sampling manifolds ${\cal M}_{i,0}$ into ${\cal M}$, 
\begin{align} 
	\int\limits_{{\cal M}_{i,0}} d\mu_{i,0} \, \rho_{i,0}(\phi) \,{\cal O}(\phi) = \int\limits_{{\cal M}} d\mu \, \rho_{i,0}(\phi) \,{\cal O}(\phi) \,,
\label{eq:Cauchyrhoi}
\end{align}
where $d\mu_{i,t}$ are the tangential measures along the manifolds ${\cal M}_{i,t}$ and $d\mu=d\phi$ when ${\cal M}$ is assumed to be flat. This allows us to compute expectation values with the distribution $p_0=p^{(w)}$ and the real sampling manifold ${\cal M}$, 
\begin{subequations} 
\label{eq:ComplexPIKwoSignProblemPre}
\begin{align} \nonumber 
		\int\limits_{{\cal M}}  d\mu\, p_0(\phi) \,{\cal O}(\phi)&\\[1ex] 
		&\hspace{-2cm}= \sum_{i=1}^N w_i \int\limits_{{\cal M}}  d\mu\, \rho_{-1}(\phi) {\cal O}\bigl(\phi_{i,0}(\phi)\bigr)\,,
		\label{eq:EquivalenceSamplingPre}
\end{align} 
with
\begin{align} 
		\phi_{i,t}(\phi) = \phi + \int_{-1}^{t} ds \, \dot \phi_{i,s} (\phi_s)\,,\qquad t\in[-1,0]\,. 
		\label{eq:IntegrateMapPre}
\end{align}
\end{subequations} 
Expectation values with $p^{(w)}(\phi)$ on the real sampling manifold ${\cal M}_{-1}$ are weighted sums of expectation values with the $\rho_i$, for which we can solve the respective sign problem. Since all weights are positive, $w_i\in\mathbb{R}^+$, this decomposition does not reintroduce a sign problem at the level of the sum. This is in contrast to Lefschetz thimbles: there, each thimble contributes with a complex phase, and the interference between these phases may constitute a residual sign problem, see e.g.~\cite{Fujii:2013sra,Cristoforetti:2014gsa,Berger:2019odf,Aarts:2026uiu}. Moreover, the single distributions $\rho_i$ typically come with multiple, often infinitely many, thimbles themselves, see e.g.~\cite{Tanizaki:2015rda}, whereas each of them is sampled with a single real PIKfold. Hence, generically the number of contributing thimbles grows qualitatively faster than the number of distributions $\rho_i$. In summary, the fan of PIKs solves the sign problem of $p_0(\phi)= p^{(w)}(\phi)$.

Now we use that $(p,{\cal M})$ is in the equivalence class of $(p^{(w)},{\cal M})$, which entails that there is a PIK $\dot\phi_t^{(w)}$ with $t\in [0,1)$ that maps $p^{(w)}$ into $p$, see \labelcref{eq:PIKw} and \Cref{fig:ComplexRealPIKfan}. 
This allows us to define PIKs from $t=-1$ to $t=1$ as a combination of the PIKs $\dot \phi_{i,t}$ for $t\in[-1,0)$ with $\dot\phi_t^{(w)}$ for $t\in[0,1)$, to wit 
\begin{align} 
\dot \phi^{(w)}_{i,t} = \dot \phi_{i,t}\,\theta(-t ) + \dot \phi^{(w)}_{t}\,\theta(t ) \,,
\label{eq:FullPIK} 
\end{align} 
with the Heaviside $\theta$-function, see \Cref{fig:ComplexRealPIKfan}. This defines a combination of PIKs that fan out from a real normal distribution $\rho_\textrm{real}$ with accompanying PIKfolds ${\cal P}_i$ with $t\in[-1,1]$ for the sampling pairs 
\begin{subequations} 
	\label{eq:PIKfan}
\begin{align} 
	(\rho_{i,-1}\,,\,{\cal M}) \stackrel{\dot\phi^{(w)}_{i,t}}{\longrightarrow} 	(\rho_{i,1}\,,\,{\cal M}_{i,1})\,,
\label{eq:RealPIK-i}
\end{align} 	
with 
\begin{align} 
 \rho_{i,-1}=\rho_\textrm{real}\,,\quad {\cal M}_{i,-1}={\cal M}\,, 
	\label{eq:RealPIK-i-initial}
\end{align} 	
and the sampling identity for all $t\in[-1,1]$,
\begin{align} 
\int\limits_{{\cal M}_{i,t}} \, d\mu_{i,t} \,\rho_{i,t} \,{\cal O}(\phi) = 	\int\limits_{{\cal M}} \, d\mu \,\rho_{-1} \,{\cal O}\bigl(\phi^{(w)}_{i,t}(\phi)\bigr)\,,
\label{eq:EquivSamplingrhoi}
\end{align} 
with the fields 
\begin{align} 
	\phi^{(w)}_{i,t}(\phi) = \phi + \int_{-1}^{t} ds \, \dot \phi^{(w)}_{i,s} (\phi_s)\,. 
	\label{eq:IntegrateMap}
\end{align}
\end{subequations} 
We call this fan of PIKs with the property \labelcref{eq:EquivSamplingrhoi} a \textit{PIKfan}, and the individual sampling pairs $(\rho_{i,t},{\cal M}_{i,t})$ its \textit{ribs}, see \Cref{fig:ComplexRealPIKfan}. Importantly, the sampling pairs $(\rho_{i,-1}\,,\,{\cal M})$ are real and the $\rho_{i,t}$ satisfy the Wegner equation with $\dot\phi^{(w)}_{i,t}$ by construction. Hence, we have real PIKfolds 
\begin{align} 
{\cal P}_i=\{(\rho_{i,t}, {\cal M}_{i,t})\,| \, t\in[-1,1]\}\,.
\label{eq:RealPi}
\end{align} 
With the initial assumption that $(p,{\cal M})$ is in the equivalence class ${\cal E}(p^{(w)},{\cal M})$, the final sampling prescription follows readily as  
\begin{align} 
	\int\limits_{{\cal M}}  d\mu \, p(\phi) \,{\cal O}(\phi)= \sum_{i=1}^N w_i  \int\limits_{{\cal M}}  d\mu\, \rho_{-1}(\phi) \,{\cal O}\bigl(\varphi^{(w)}_i(\phi)\bigr)\hspace{.01cm},
	\label{eq:EquivalenceSampling}
\end{align} 
with the push-forward configurations \labelcref{eq:IntegrateMap} for $t=1$, 
\begin{align} 
	 \varphi_i^{(w)}(\phi)=\phi_{i,1}^{(w)}(\phi)\,.
	\label{eq:IntegrateMap2}
\end{align}
Note that we still have to verify the existence of $(p,{\cal M})\in {\cal E}(p^{(w)},{\cal M})$. 
This is discussed in detail in \Cref{sec:PropertiesOfPIKfans,sec:FindingPIKfans}; for now we simply state that the tools to verify this existence are naturally provided in the construction of the PIKfan itself. This completes our construction. 

The practical sampling procedure is done by drawing a set of $N_s$ configurations ${\cal S}_{-1}=\{\phi_n\}$ with the distribution $\rho_{-1}=\rho_\textrm{real}$ on ${\cal M}$. The expectation values of operators ${\cal O}$ with the target distribution $p(\varphi)$ are then provided by 
\begin{align}
	\langle \mathcal{O}\rangle_{(p, \mathcal{M})} \simeq \frac{1}{N_s} \sum_{i=1}^N w_i \sum_{\phi_n\in{\cal S}_{-1}} \mathcal{O}\left(\varphi^{(w)}_i(\phi_n)\right) \,. 
	\label{eq:SampleOt}
\end{align}
Note that the $\varphi^{(w)}_i(\phi_n)$ are nothing but the push-forward samples, and the push-forward is the heart of the construction anyway. 
In short, we can directly consider samples on the respective manifolds $\mathcal{M}_{i,1}$ for the complex rib distributions $\rho_{i,1}$, and evaluate the expectation value via
\begin{align}
	\langle \mathcal{O}\rangle_{(p, \mathcal{M})} \simeq \frac{1}{N_s} \,\sum_{i=1}^N w_i\,\sum_{\phi_{i,n}\in{\cal S}_{i,1}}  \mathcal{O}\left(\phi_{i,n}\right) \,,
	\label{eq:SampleOtpM1}
\end{align}
where ${\cal S}_{i,1}$ is the set of configurations on ${\cal M}_{i,1}$. 

We conclude this Section with a discussion of the restriction to positive weights. On the technical side, allowing for negative weights may introduce a residual sign problem at the level of the sum, and the computational efficiency of the PIKfan architecture is potentially lost. We now argue that this is not a restriction on the theories that can be treated.

To begin with, a given sampling pair admits many decompositions, and the signs of the weights are not a property of its equivalence class. Indeed, the equivalence classes \labelcref{eq:WeightedSumDistribution2} also include sums of distributions with $w_i\in \mathbb{R}$. This is readily seen by deforming $\rho_{-1}$ with a PIK into a sum of Gau\ss ians with some negative weights, 
\begin{align} 
		(\rho_{i,-1}\,,\,{\cal M}) \stackrel{ \dot\phi_{i,t}}{\longrightarrow} 	(\rho_{i,0}\,,\,{\cal M}_{i,0})\,,
		\label{eq:+-PIK-i}
	\end{align} 	
with 
\begin{align} 
\rho_{i,0}=\sum_j v_{ji}\, \rho_{ji,0}\,, \qquad v_{ji} \in\mathbb{R}\,,\qquad \sum_j v_{ji}=1\,.   
\label{eq:rhoi0+-}
\end{align} 	
Here, the distributions $\rho_{ji,0}$ are complex Gau\ss ians obtained by rotating $\rho_{-1}$ with a respective PIK. This example already illustrates that the occurrence of negative weights in a specific decomposition is not a good characterisation of equivalence classes of sampling pairs.

Decompositions with positive weights, in turn, are guaranteed by the statistical interpretation of the theory. For a quantum system with a Hermitian Hamiltonian $\hat H$, here including the chemical potential term, the generating function has the spectral representation
\begin{align}
	Z=\mathrm{tr}\,e^{-\beta \hat H}=\sum_n e^{-\beta E_n}\,,
	\label{eq:SpectralZ}
\end{align}
with real energies $E_n$ and hence positive Boltzmann factors. In relativistic quantum field theories, the same property follows from unitarity in the form of reflection positivity, which guarantees a positive transfer matrix. Each term in \labelcref{eq:SpectralZ} defines a contribution to the path-integral distribution with the positive relative weight $w_n=e^{-\beta E_n}/Z$. A respective path-integral representation is readily derived from the sum in \labelcref{eq:SpectralZ} by reformulating it in terms of the density of states approach. 

Hence, all theories with a statistical interpretation admit decompositions with positive weights, even though a given representation, such as the expansion of the fermion determinant on a lattice, may come with weights of both signs. For related discussions, see e.g.~\cite{Salcedo:2007ji}. The one-site Hubbard model in \Cref{sec:OneSiteHubbard} illustrates this: the weights \labelcref{eq:WeightsHubbard} of its decomposition are the Boltzmann factors of the Fock states in \labelcref{eq:HubbardA}. We shall come back to the Hubbard model beyond one site at the end of \Cref{sec:PIKfanExample}.   
	
%%%%%%%%%%%%%%%%%%%%%%%%%%%%%%%
\subsection{Properties of PIKfans}
\label{sec:PropertiesOfPIKfans}

With the PIKfan we have converted the task of sampling from the complex distribution $p(\varphi)$ on ${\cal M}$ without a real PIKfold into the task of sampling from a sum of real sampling pairs, hence solving the sign problem for these types of theories. Importantly, the construction requires the existence of $(p,{\cal M})\in {\cal E}(p^{(w)},{\cal M})$. 
We shall see below that this task is resolved constructively yet again, by using the key property of the PIK architecture: configurations are pushed forward on the PIKfold and these points (and their statistical weights) are directly accessed in the computation. 
We emphasise that it is the constructive nature of this task which sets the PIK architecture apart from e.g.~Lefschetz thimbles or CLEs, where either all contributing thimbles have to be found or one has to chart out the drift force in the complexified manifold. 

To isolate the non-trivial steps in the sampling architecture with PIKfans, we first note that the step from $t=-1$ to 0 is trivial by construction, see \labelcref{eq:ComplexPIKwoSignProblemPre}. Sampling on a PIKfan with each sampling pair $(\rho_{i,t},{\cal M}_{i,t})$ is given by that on the sampling pair $(\rho_{-1},{\cal M})$, see \labelcref{eq:EquivSamplingrhoi}. This follows from the existence of the PIKs $\dot\phi_{i,t}$ for all $t\in[-1,0)$ and hence \labelcref{eq:EquivalenceSamplingPre} is obtained. 

\begin{figure}[t]
	\centering
	\includegraphics[width=0.49\textwidth]{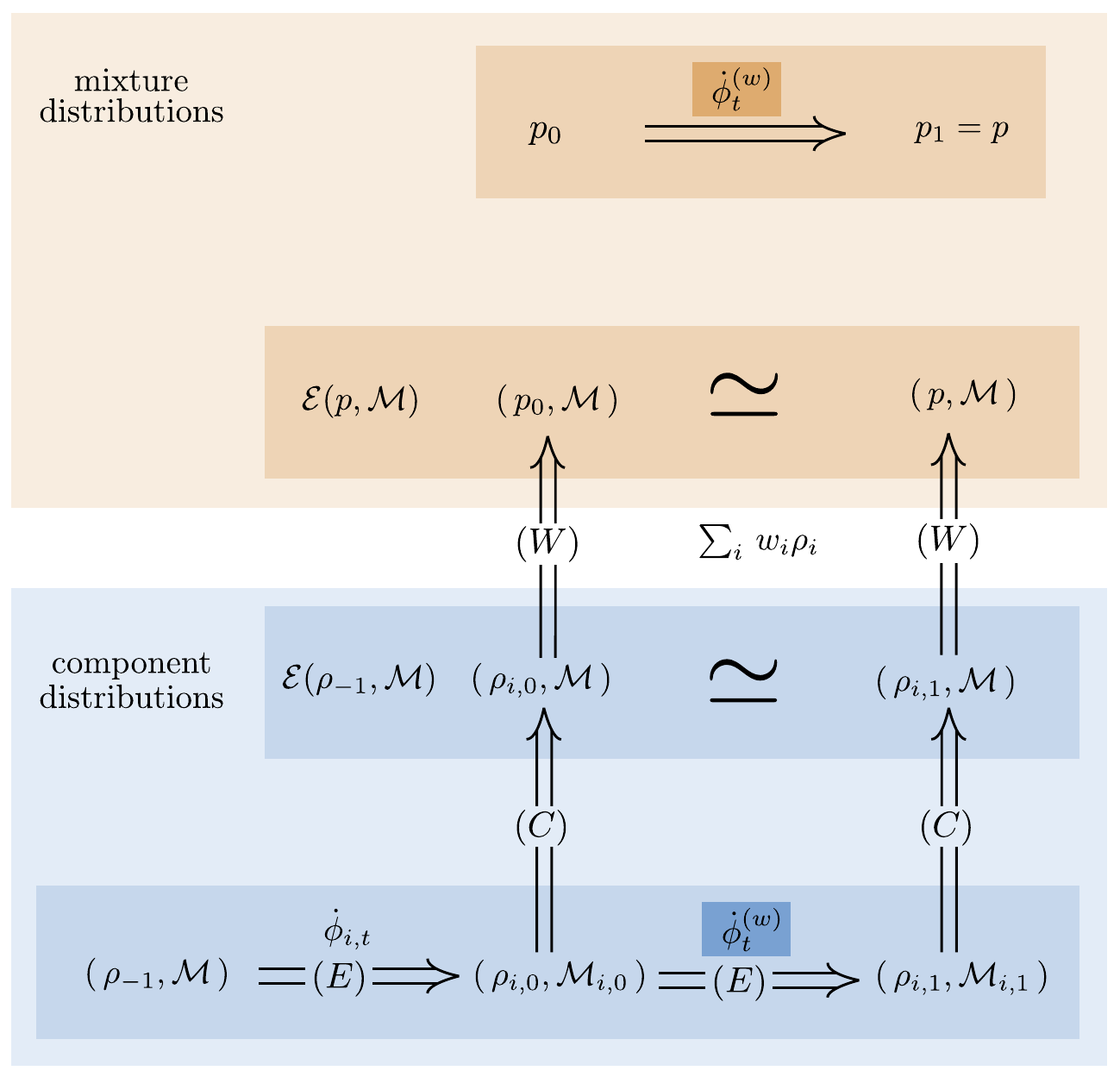}
	\caption{Equivalence-class structure underlying the PIKfan computation of observables with respect to $(p,\mathcal{M})$, see \labelcref{eq:E+C}.
	Orange (mixture distributions): the PIK $\dot\phi_t^{(w)}$ maps the mixture $p_0=p^{(w)}$ to the target $p_1=p$, when $(p_0,\mathcal{M})\simeq(p,\mathcal{M})$ lie in the same equivalence class $\mathcal{E}(p,\mathcal{M})$, see \labelcref{eq:Equivppw,eq:PIKw}.
	Blue (component distributions): the individual PIKs $\dot\phi_{i,t}$ map the real normal sampling pair $(\rho_{-1},\mathcal{M})$ to the rib pairs $(\rho_{i,0},\mathcal{M}_{i,0})$, which the common PIK $\dot\phi_t^{(w)}$ maps to $(\rho_{i,1},\mathcal{M}_{i,1})$. Both steps require the existence $(E)$ of the flows, and returning from $\mathcal{M}_{i,0}$ and $\mathcal{M}_{i,1}$ to $\mathcal{M}$ requires Cauchy's theorem $(C)$. When these conditions are met, all rib pairs lie in $\mathcal{E}(\rho_{-1},\mathcal{M})$.
	Then, the Wegner equation $(W)$ links the two levels: the weighted sums $\sum_i w_i\rho_i$ give $(p_0,\mathcal{M})$ and $(p,\mathcal{M})$.
	\hspace*{\fill}}
	\label{fig:SketchECW}
\end{figure}

The evolution of the weighted sum $p_0=p^{(w)}$ at $t=0$ to the target distribution $p_1=p$ is done by using the two non-trivial properties denoted symbolically by $(E)$ and $(C)$.
Here, $(E)$ stands for \textit{Existence} of the manifolds ${\cal M}_{i,1}$, that is, the existence and analyticity of $\dot\phi^{(w)}_t$ on the manifolds ${\cal M}_{i,t}$ for all $t\in [0,1)$. The property $(C)$ stands for the applicability of \textit{Cauchy's} theorem to deform the manifolds, that is, the absence of non-analyticities over the deformation. The complete  architecture and the use of $(E,C)$ and the Wegner equation $(W)$ in the different steps are illustrated in \Cref{fig:SketchECW}. 

We start our analysis with the right-hand side of \labelcref{eq:SampleOt}, which defines our real sampling procedure. This is the sampling on the fan of distributions $\rho_{i,1}$ and it only requires that the samples $\phi\in {\cal S}_{-1}$ can smoothly be pushed forward to $\varphi^{(w)}_i(\phi)\in{\cal S}_{i,1}$, that is, the existence of the PIK $\dot\phi^{(w)}_t$ on the manifolds ${\cal M}_{i,t}$, which also ensures the absence of ripping and hence the existence of the PIKfolds ${\cal P}_i$ defined in \labelcref{eq:RealPi}. This is nothing but the property $(E)$, i.e., with $(E)$ satisfied, the PIKfan expectation value is defined as
\begin{subequations} 
	\label{eq:E+C}
\begin{align} \nonumber 
	\langle \mathcal{O} \rangle^{\phantom{1}}_{\textrm{fan}}\overset{(E)}{:=}&\, \sum_{i=1}^N w_i \int\limits_{\mathcal{M}} d\mu\, \rho_{-1}(\phi) \,\mathcal{O}\left(\varphi_i^{(w)}(\phi)\right) \\[1ex] 
	 =&\, \sum_{i=1}^N w_i \int\limits_{\mathcal{M}_{i,1}} d\mu_{i,1} \, \rho_{i,1}(\phi) \,\mathcal{O}\left(\,\phi \,\right) \,.
\label{eq:SamplingPrePIKfan}
\end{align} 
\Cref{eq:SamplingPrePIKfan}, the sampling on the PIKfan, is identical to the sampling with $(p,{\cal M})$ if the sampling with $\rho_{i,1}$ on the manifolds ${\cal M}_{i,1}$ can be deformed into sampling with $\rho_{i,1}$ on ${\cal M}$ with Cauchy's theorem. This is nothing but the property $(C)$, i.e., with $(C)$ satisfied, the PIKfan expectation value agrees with the physical one:
\begin{align}\nonumber
\langle \mathcal{O} \rangle^{\phantom{1}}_{\textrm{fan}}	
	&\overset{(C)}{=} \sum_{i=1}^N w_i \int\limits_{\mathcal{M}} d\mu \, \rho_{i,1}(\phi) \,\mathcal{O}\left(\,\phi \,\right) \\[1ex]
	&\overset{(W)}{=} \int\limits_{\mathcal{M}} d\mu \, p(\varphi) \,\mathcal{O}\left(\,\varphi \,\right) = \langle \mathcal{O} \rangle^{\phantom{1}}_{\textrm{physics}} \,, 
	\label{eq:Cauchy}
\end{align}
\end{subequations}
where $(W)$ stands for the Wegner equation, which is nothing but the PIK architecture itself.

\begin{figure}[t]
	\centering
	\begin{subfigure}{.49\linewidth}
		\centering
		\includegraphics[width=\linewidth]{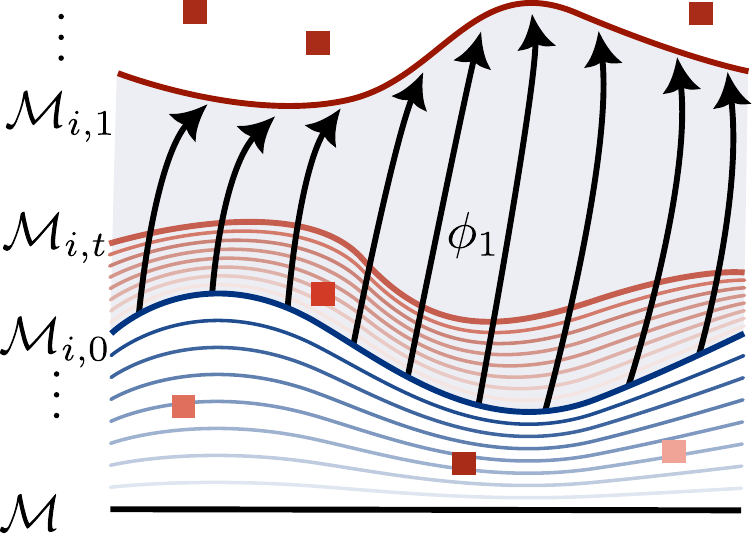}
		\subcaption{Illustration of $(E)$.}
		\label{fig:E_concept_working}
	\end{subfigure}
	\begin{subfigure}{.49\linewidth}
		\centering
		\includegraphics[width=\linewidth]{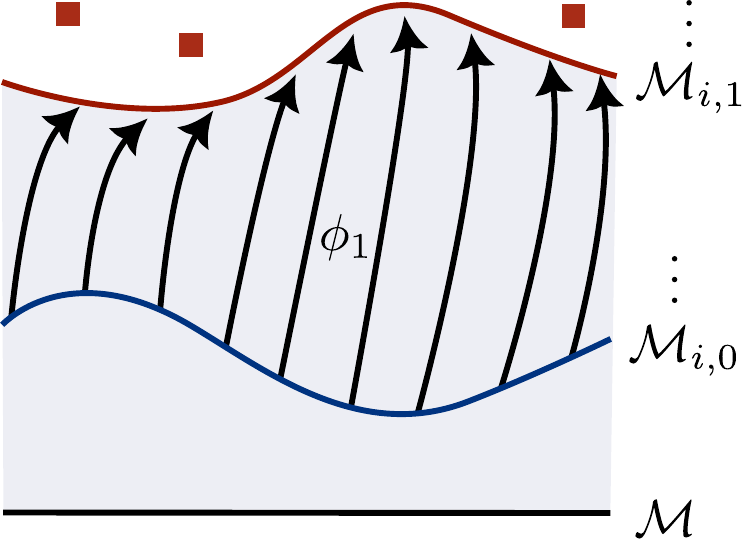}
		\subcaption{Illustration of $(C)$.}
		\label{fig:C_concept_working}
	\end{subfigure}
	\caption{Illustration of the conditions $(E)$ and $(C)$ for a zero-dimensional example. The manifolds $\mathcal{M}$ (black), $\mathcal{M}_{i,0}$ (dark blue), $\mathcal{M}_{i,t}$ (light red), and $\mathcal{M}_{i,1}$ (dark red) are indicated by lines. Different shades of the same colour indicate the manifolds at different times (darker means later). The map $\phi_1$ is shown with black arrows and denotes the transport map from $t=0$ to $t=1$. The singularities of this map are shown by red squares. Their translucency  indicates the time at which they appear and corresponds to the shade of the manifolds. The shaded area in $(b)$ shows the region where $\phi_1$ must be free of singularities. \hspace*{\fill}}
	\label{fig:Illustration_Nontrivial_Steps}
\end{figure}

We proceed with investigating how the properties $(E)$ and $(C)$ in \labelcref{eq:E+C} can be satisfied in the practical construction. 
In \Cref{fig:Illustration_Nontrivial_Steps}, the properties $(E)$ and $(C)$ and their potential failures are illustrated for a zero-dimensional example. Here, $\mathcal{M}$ is a black line at the bottom and $\mathcal{M}_{i,t}$ with $t\in(-1,0]$ are blue lines with decreasing translucency. The manifolds ${\cal M}_{i,0}$ are indicated by a dark blue line. The manifolds $\mathcal{M}_{i,t}$ with $t\in(0,1]$ are indicated with red lines with decreasing translucency, and $\mathcal{M}_{i,1}$ is indicated by 
a dark red line. Moreover, the map $\phi_1$, the transport map from $t=0$ to $t=1$, is illustrated by black arrows. The singularities of this map are shown by red squares. Their translucency indicates the RG-time at which they form and corresponds to the translucency of the manifolds. 

\Cref{eq:E+C} summarises three steps with the properties $(E)$, $(C)$ and the Wegner equation $(W)$. While the last step with $(W)$ is trivial by construction, the steps with $(E)$, $(C)$ require the absence of non-analyticities in the evolution of the sampling manifolds with $\dot\phi_{i,t}$ and on the final PIKfolds ${\cal P}_i$.  We reiterate $(E)$ and $(C)$ with the schematic illustration as well as $(W)$ as follows:
\begin{itemize}
  \item{$(E)$:} \Cref{eq:SamplingPrePIKfan}, that is sampling with the PIKfan, requires that the manifolds ${\cal M}_{i,t}$ do not rip and hence ${\cal M}_{i,1}$ exists. This no-ripping condition is satisfied if $\dot \phi^{(w)}_t$ is analytic on all ${\cal M}_{i,t}$ for $t\in [0,1)$. 
    This is illustrated in \Cref{fig:E_concept_working}. Here, the moving manifold $\mathcal{M}_{i,t}$ (light red) does not encounter singularities while evolving from $\mathcal{M}$ (black) to $\mathcal{M}_{i,1}$ (dark red). 
    Singularities either appear away from $\mathcal{M}_{i,1}$ (above the dark red line in the illustration) or only appear in regions already swept by the manifold at earlier times, that is below $\mathcal{M}_{i,t}$ in \Cref{fig:E_concept_working}.

  \item{$(C)$:} \Cref{eq:Cauchy} requires Cauchy's theorem to hold. It can be applied in the absence of non-analyticities in the maps $\varphi^{(w)}_i(\phi)$ on the real PIKfolds ${\cal P}_{i}$ defined in \labelcref{eq:RealPi}. 
This is illustrated in \Cref{fig:C_concept_working}. The integrated transport maps $\varphi^{(w)}_i(\phi)$ and $\rho_{i,1}$ must be free of singularities in the shaded area swept over when deforming ${\cal M}_{i,1}$ into ${\cal M}$.

  \item{$(W)$:} The second step in \labelcref{eq:Cauchy} follows from the Wegner equation. The $\rho_{i,0}$ are mapped with the PIK $\dot\phi^{(w)}_t$ to $\rho_{i,1}$. This is the PIK which maps $p_0=p^{(w)}$ to $p_1=p$.
\end{itemize}
Importantly, tracking down the location of these singularities or their absence in the map $\dot \phi^{(w)}_t$ on the real PIKfolds ${\cal P}_{i}$ does not come at the price of searching for singular (or special) points in the complexified sampling manifold. It is part of the computation and hence avoids the computationally hard part of other methods based on the CLEs and the Lefschetz thimbles. We also emphasise that the absence of non-analyticities on the real PIKfolds ${\cal P}_i$ 
\textit{guarantees} that $(E)$ and $(C)$ hold but does not cover the general case. For example, 
$(C)$ also holds if the contributions from potential singularities cancel out. This is but one of the possible generalisations; more details are provided in \Cref{sec:NoRip,sec:PIKfanCauchy}.

%%%%%%%%%%%%%%%%%%%%%%%%%%%%%
\subsection{Finding PIKfans}
\label{sec:FindingPIKfans}

This leaves us with the task of finding PIKfans. As discussed above and illustrated further in \Cref{sec:OneSiteHubbard}, for a given fermionic distribution $p=p_1$, there exist initial distributions $p_0=p^{(w)}$ for which $(E)$ and $(C)$ are satisfied, and others for which no regular maps to $p_1$ can be found. Finding PIKfans thus means choosing the weights $w_i$ and rib distributions $\rho_{i,0}$ with $p^{(w)} = \sum_i w_i \rho_{i,0}$ such that the common PIK $\dot\phi^{(w)}_t$ satisfies $(E)$ and $(C)$.

To begin with, a regular flow can neither create nor destroy zeros of the distribution, see \Cref{sec:zeros_and_transport} for more details. Hence, $(E)$ is in particular violated if the path requires a zero to be created in a region that carries weight, as in \Cref{fig:individual_zero_rip}, where $p_0$ has no zeros while $p_t$ does.

Na\"ively, searching for the zero hypersurfaces ${\cal Z}_t$ in \labelcref{eq:ptZero} in high-dimensional complex spaces is a prohibitively difficult task. However, PIKs encounter the intersection sets ${\cal Z}_t\cap{\cal M}_t$ during their construction. Hence, a violation of $(E)$ is directly reflected in the map's regularity, which can for instance be probed during the construction of the PIK by indicators like $\Lambda_t$ defined in \labelcref{eq:regularity_indicator} evaluated on the samples of all ribs.
For example, $\Lambda_t$ diverges at the time $t_\textrm{rip}$ at which a zero has to be created, see e.g.~\Cref{fig:individual_zero_rip,fig:E_problem_observable}, and the samples on which it diverges indicate this zero. As the samples carry the weight determined by $w_i$ and $\rho_{-1}$, only zeros that are reached with a relevant weight show up, and \textit{relevant} is thus directly
defined by the notion of the initial weights. In short, one uncovers the relevant complex structure of $p_t$ as one constructs the PIKs. The diagnostic specifies what is missing in $p_0$. 
For instance, the (onset of a) diverging $\Lambda_t$ states that either $p_0$ or the path must carry a zero where the flow had to remove or create one.

The same structure underlies $(C)$. Both conditions are statements about the regularity of the flow: $(E)$ requires it on the manifolds $\mathcal{M}_{i,t}$, while $(C)$ requires it in the region swept when deforming $\mathcal{M}_{i,t}$ into $\mathcal{M}$, where the maps $\phi^{(w)}_{i,t}$ and distributions $\rho_{i,t}$ must neither develop branch points nor divergences. The tools are therefore the same: the analytic structure of $\rho_{i,t}$ is known, as they evolve from the trial sampling pair $(\rho_{-1},{\cal M})$. Here it pays off that we have restricted ourselves to ${\cal M}_{-1}={\cal M}$. The kernel $\dot\phi^{(w)}_t$ is accessible, and the relevant region is bounded by $\mathcal{M}$ and the manifolds $\mathcal{M}_{i,t}$ computed during the flow. Probes of the regularity of the flow such as $\Lambda_t$ and versions thereof, extended from the samples to this region, can thus be used to track potential obstructions of $(C)$. Moreover, as for $(E)$, no search for zeros is required: the zeros of $p_0$ are known since $p_0$ is chosen, and their transport along the flow follows from \labelcref{eq:field_transport_zero_transport_app,eq:zeros_along_transport} with the computed kernel $\dot\phi^{(w)}_t$. This is illustrated in \Cref{sec:PIKfanCauchy}, where following the zeros of $p_0$ along the flow reveals a violation of $(C)$ that would not have been detected otherwise, while a different path $p_t$ satisfies $(C)$.

Thus, finding a PIKfan instantiates an optimisation task to find $w_i$, $\rho_{i,0}$ and $p_t$ that is governed by the information generated within the PIK construction itself. This sets it apart from other approaches, which require an additional search for zeros, special points or auxiliary regularisation conditions. A precise algorithmic implementation will be provided in a forthcoming work.

%%%%%%%%%%%%%%%%%%%%%%%%%%%%%%%%%%%%%%
\section{One-site Hubbard model and beyond}
\label{sec:OneSiteHubbard}

In this Section we consider applications of the PIKfan architecture to theories consisting of a sum of complex  Gau\ss ians. This also includes the Hubbard model, which suffers from a sign problem similar to that of QCD at finite density. For specific approaches to tackle the sign problem in the Hubbard model, see e.g.~\cite{Tanizaki:2015rda,Fukuma:2019wbv,Berger:2019odf,Gantgen:2023byf,Ulybyshev:2024kdr,Schuh:2026dvp,Temmen:2026ydp}. Like the conserved quark number in QCD, the conserved fermion number in the Hubbard model can be coupled to the chemical potential $\mu$, leading to a phase in the distribution.  In the atomic limit, fermions can no longer hop to neighbouring sites, and the model Hamiltonian is decomposed into the one-site Hamiltonian at each site. 

This simple example is not in the equivalence class of real PIKfolds for all chemical potentials, and we demonstrate the ripping of the manifold ${\cal M}_t$ in \Cref{sec:PolesAndPiks}. The ripping of the integration contour is connected to zeros of the distribution, which are subsequently treated with a PIKfan in \Cref{sec:PIKfanExample}.
Finally, in \Cref{sec:phi4_mixture}, we show that this approach generalises beyond the one-site Hubbard model: while the latter can be written as a simple mixture of Gau\ss ians after a Hubbard--Stratonovich transformation, see \labelcref{eq:HubbardZ}, we also consider complex mixtures of $\phi^4$ theories.

%%%%%%%%%%%%%%%%
\subsection{Basics}
\label{sec:1SiteHubbard}

Using the number operators for spin-up and spin-down fermions, $\hat{n}_\uparrow$ and $\hat{n}_\downarrow$, the one-site Hamiltonian reads
\begin{align}
  \hat{H} = U \hat{n}_\uparrow \hat{n}_\downarrow - \mu
  (\hat{n}_\uparrow + \hat{n}_\downarrow) \,.
\label{eq:Hamiltonian}
\end{align}
With this simple Hamiltonian the generating function can be computed analytically. Still, this example captures the essence of the sign problem: direct simulations with importance sampling fail due to the sign problem, which is discussed in \Cref{sec:PolesAndPiks}. 

In the operator formalism, it is a straightforward task to compute the partition function using the Fock state basis.  Because $\hat{n}_{\uparrow,\downarrow}$ take the eigenvalues $0$ or $1$, there are four eigenvalues of $\hat{H}$, namely, $0$, $-\mu$, $-\mu$, $U-2\mu$.  Therefore, the partition function is
\begin{align}
  \mathrm{tr}\bigl[ e^{-\beta\hat{H}} \bigr]
  = 1 + 2\, e^{\beta\mu} + e^{\beta(2\mu - U)} \,,
  \label{eq:HubbardA}
\end{align}
from which we can compute thermodynamic quantities. 
A path-integral representation of the normalised generating function $Z=\int d\phi\,p(\phi)=1$ is provided with the distribution 
\begin{subequations}
	\label{eq:HubbardModel}
\begin{align}
	p(\phi)= e^{-S(\phi)}\,,\qquad S(\phi)= \hat S(\phi) +\ln {\cal N}\,. 
	\label{eq:HubbardZ}
\end{align}
Here, we deviate from~\cite{Ihssen:2026njd} and define the field-dependent part of the action not via $\hat S(\phi)=S(\phi) -S(0)$ but as
\begin{align}
	\hat S(\phi) = \frac{\beta}{2U} \phi^2 - 
	2 \ln \left[ 1 + e^{ \beta \left( i \phi + \mu + U/2\right)} \right]\,,
\label{eq:HubbardhatS}
\end{align}
in order to match with conventions from the literature, see e.g.~\cite{Tanizaki:2015rda}. With this, the normalisation reads 
\begin{align} 
	\mathcal{N} = \sqrt{\frac{2\pi U}{\beta}} \left[ 1 + 2 e^{\beta \mu} + e^{\beta(2\mu - U)} \right]
\label{eq:NormalisationHubbard} 
\end{align} 
which guarantees $Z=1$. 
\end{subequations}
For a detailed derivation, see for instance~\cite{Tanizaki:2015rda}. The second term on the right-hand side of \labelcref{eq:HubbardhatS} is a counterpart of the log of the Dirac determinant in QCD at finite density. We emphasise that the similarity is due to the fermionic nature, and in this sense the one-site Hubbard model can be regarded as a well-motivated toy model to investigate the sign problem in QCD at finite density.

%%%%%%%%%%%%%%%%%
\subsection{A severe sign problem}
\label{sec:PolesAndPiks}

\begin{figure}[t]
	\centering
	\includegraphics[width=\linewidth]{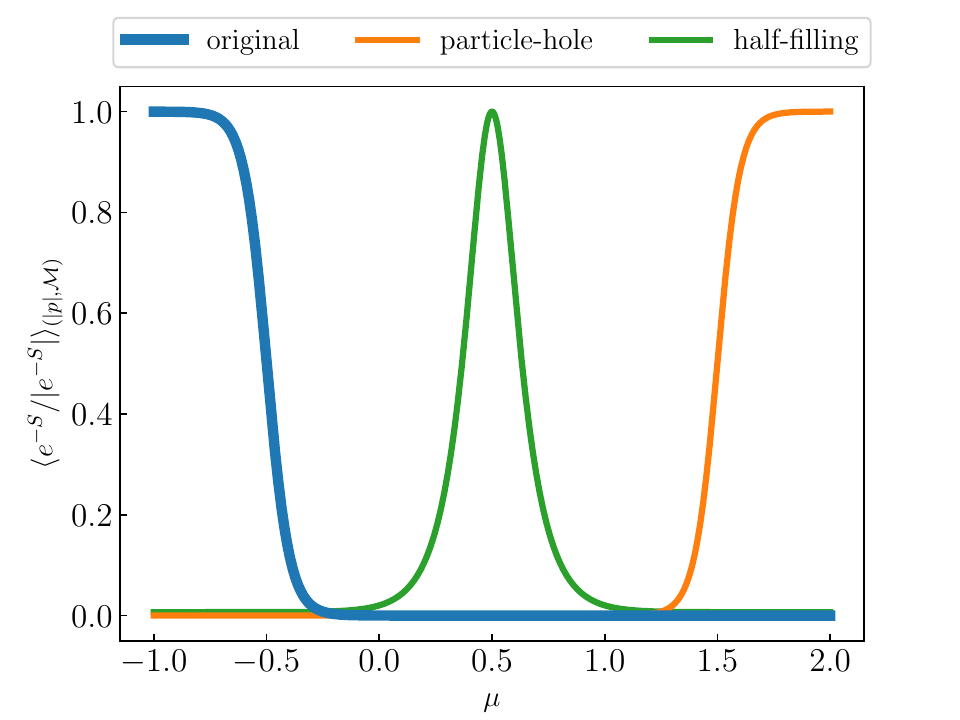}
	\caption{Average sign $\langle e^{- \hat S}/|e^{-  \hat S}|\rangle_{(|p|,{\cal M})}$ in 	\labelcref{eq:AverageSign} of the one-site Hubbard model for various chemical potentials $\mu$ while keeping $U = 1$ and $\beta = 10$ fixed. The average sign is shown for different variable transformations.  For more details, see \Cref{sec:variable_transformations_hubbard}. \hspace*{\fill}}
	\label{fig:average_sign_hubbard}
\end{figure}

One might think that the one-site Hubbard model is too simple to show a sign problem, let alone a severe one similar to that in QCD at finite density. However, once the fermionic determinant is included in the action, we see that 
\labelcref{eq:HubbardZ} is a good proxy for the QCD sign problem: the action is complex-valued, and observables cannot be computed na\"ively using importance sampling methods for some chemical potential range. The severity of this problem is captured by the average sign 
\begin{align} 
	\left\langle \frac{e^{-  \hat S}}{\left| e^{- \hat S}\right|}\right\rangle_{(|p|,{\cal M})}=\frac{\int\limits_{\cal M} d\phi\,\Big|e^{-\hat S(\phi)}\Big| \frac{e^{- \hat S}}{\left|e^{- \hat S}\right|}}{\int\limits_{\cal M} d\phi\,\Big|e^{-\hat S(\phi)}\Big|}=\frac{\mathcal{N}_p}{\mathcal{N}_{|p|}}\,,
	\label{eq:AverageSign}
\end{align}
where $\mathcal{N}_p$ and $\mathcal{N}_{|p|}$ are the normalisation constants of the physical and the phase-quenched distribution, respectively, with $\mathcal{N}_p=\mathcal{N}$ in \labelcref{eq:NormalisationHubbard}. This observable quantifies how well general observables of the theory can be computed by importance sampling with respect to the phase-quenched distribution $|p|$. For the present case, this is shown in \Cref{fig:average_sign_hubbard} for different variable transformations of \labelcref{eq:HubbardZ}; details are provided in \Cref{sec:variable_transformations_hubbard}. While some transformations significantly improve the average sign, we observe that they cannot restore it for all parameter sets, disallowing sampling from the phase-quenched distribution for practical purposes in these cases.

Moreover, note that the action in \labelcref{eq:HubbardZ} has some characteristic points $\phi^{\mathrm{(s)}}_n$, which we can identify as the zeros of the distribution $p(\phi)$,  
\begin{align}
  \phi^{\mathrm{(s)}}_{n} = (2 n + 1) \frac{\pi}{\beta} + i \left(\mu + \frac{U}{2}\right)\,.
  \label{eq:HubbardPoles}
\end{align}
At the  $\phi^{\mathrm{(s)}}_n$ the action hits a singularity. In the PIK approach the action $S(\phi)$ in \labelcref{eq:HubbardModel} is rewritten as a time-dependent path with the final physical action $S_{t=1} = S$. Consequently, the zeros, $\phi^{\mathrm{(s)}}_{n,t}$, of the distribution also move through the complex plane. If the sampling manifold ${\cal M}_t$ sweeps over one of these zeros, the PIK $\dot\phi_t(\phi^{\mathrm{(s)}}_{n,t})$ also hits a singularity. Then, the zeros \labelcref{eq:HubbardPoles} of the distribution terminate the existence of real PIKfolds, see \labelcref{eq:ptZero}. For a detailed discussion of this mechanism, see \Cref{sec:RippingExample,sec:PIKsandPoles,sec:zeros_and_transport}.

\begin{figure*}
	\begin{subfigure}{.32\linewidth}
		\includegraphics[width=\linewidth]{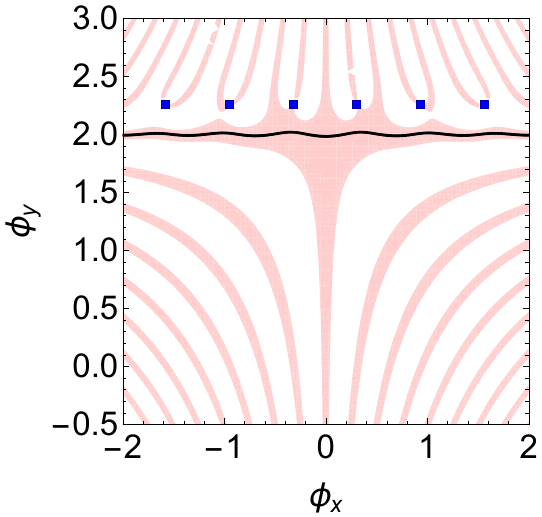}
		\subcaption{$\mu = 1.75$}
		\label{fig:pik_particle_hole_contour}
	\end{subfigure}
	\begin{subfigure}{.32\linewidth}
		\includegraphics[width=\linewidth]{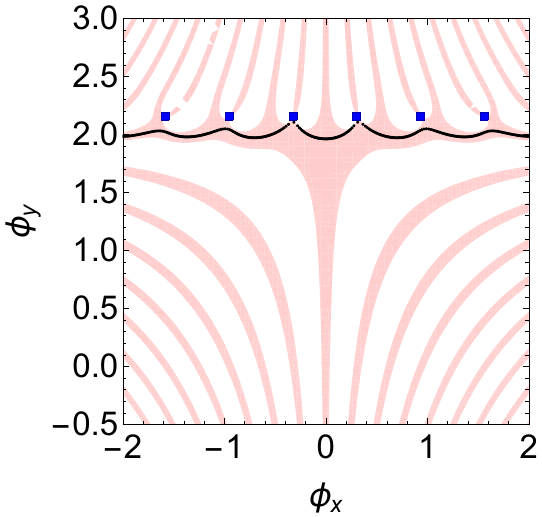}
		\subcaption{$\mu = 1.65$}
	\end{subfigure}
	\begin{subfigure}{.32\linewidth}
		\includegraphics[width=\linewidth]{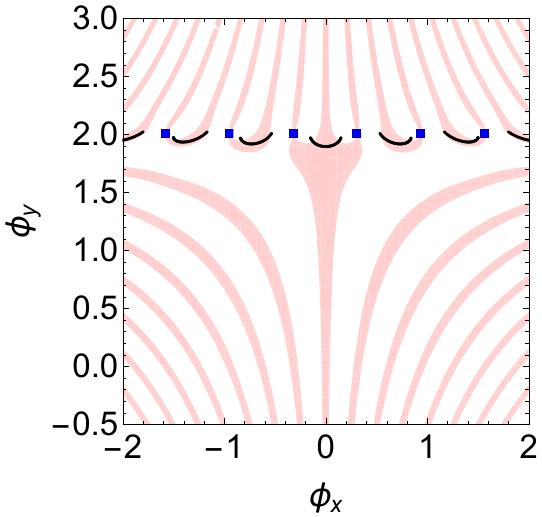}
		\subcaption{$\mu = 1.5$}
		\label{fig:FirstRIP}
	\end{subfigure}
	\centering
	\caption{Final sampling manifold in the complex plane with $\phi=\phi_x + i\phi_y$ for the single-PIKfold approach discussed in \Cref{sec:PIKsandPoles} for different chemical potentials.  The sampling manifold is indicated in \textit{black}, whereas the \textit{blue} dots indicate the zeros, \labelcref{eq:HubbardPoles1}, of the distribution $p$, see \labelcref{eq:HubbardModel}. The red areas are inserted for orientation and indicate $\mathrm{Im} \, S(\phi) \pmod{2 \pi} < \frac{\pi}{4}$ or $\mathrm{Im} \, S(\phi) \pmod{2 \pi} > \frac{7 \pi}{4}$. \hspace*{\fill}}
	\label{fig:PIKfoldPlot}
\end{figure*}
%

%%%%%%%%%%%%%%%%
\subsubsection{The ripping PIKfold}
\label{sec:PIKsandPoles}

We find that whenever a zero of the distribution approaches the PIKfold, a cut starts to develop in the manifold. However, the example of the one-site Hubbard model slightly differs from the one discussed in \Cref{sec:RippingExample}, since the PIK $\dot \phi_t$ is such that the samples are repelled by the zero. Thus, the rip shows as a jump in $\dot \phi_t$.
Due to the ripping, the existence condition $(E)$ from \Cref{sec:PropertiesOfPIKfans} is violated, as in \Cref{sec:RippingExample}, and $\mathcal{M}_t$ is no longer a valid integration manifold.
To demonstrate the above, we consider a simple path which evolves samples from a Gau\ss ian distribution at $t=0$ to the full action in \labelcref{eq:HubbardModel} at $t=1$:
\begin{subequations}
	\begin{align}\label{eq:actionPath1}
		S_t (\phi) = \frac{\beta}{2U} \phi^2 - 2 \ln\left[ 1 + e^{t \beta \left( i \phi + \mu + U/2\right)}\right] + \ln \mathcal{N}_t\,,
	\end{align}
	with time-dependent zeros of the distribution at 
	\begin{align}\label{eq:HubbardPoles1}
		\phi^{\mathrm{(s)}}_{n,t}= (2 n +1 ) \frac{\pi }{t\beta} + i \left(\mu + \frac{U}{2}\right)\,.
	\end{align}
\end{subequations}
The path \labelcref{eq:actionPath1} allows one to determine an analytical expression for $\dot \phi_t$ that solves the Wegner equation \labelcref{eq:PIKflow}. A detailed computation is provided in \Cref{app:analyticSol}. 
The PIKfold at $t=0$ is initialised on the real axis $\mathcal{M}_0 = \mathbb{R}$, where Gau\ss ian samples can be drawn. These are then evolved using \labelcref{eq:phiMap} and the corresponding $\dot \phi_t$ for the path \labelcref{eq:actionPath1}. Throughout the RG-time evolution of the distribution, its zeros remain at a constant imaginary value and only move in the direction of the real axis. Hence, any PIKfold with $\mathrm{Im}[\mathcal{M}_t] \lesssim \mu + \frac{U}{2}$ does not encounter the zeros. 

We depict the final sampling manifold for this example in \Cref{fig:PIKfoldPlot} with the parameter choices $U=1$, $\beta =10$ at various chemical potentials. The results displayed in panels (a) and (b) correspond to $\mu = 1.75$ and $\mu=1.65$ and are valid sampling manifolds as they remain far enough away from the zeros and show no ripping. In (c) at $\mu = 1.5$, the zeros have approached the manifold too closely.
The contour rips and we can no longer evaluate observables from the samples.

The complex structure of the one-site Hubbard model depicted in \Cref{fig:PIKfoldPlot} allows us to make the following observation: for chemical potentials $\mu \gtrsim 3U/2$ there exists a contour with relatively constant phase at $\mathrm{Im} [\phi] \approx 2\, U$, which is broken once the zeros pass through it, i.e., $\mu \lesssim 3U/2$.  This observation allows us to distinguish a first type of sign problem in the one-site Hubbard model: all actions with $\mu \gtrsim 3U/2$ are connected to sampling from a real Gau\ss ian by means of a single real PIKfold. The solution of the PIKfold itself resembles the variable transformation used in \cite{Tanizaki:2015rda} and that of a single simple complex Gau\ss ian.

Similarly, one finds that actions with $\mu \lesssim - U/2$ are also connected to sampling from a real Gau\ss ian by means of a single real PIKfold. This validates the assessment that the sign problem is less severe for  $ \mu  \lesssim -U/2$ and $\mu \gtrsim 3U/2$, which is commonly found in the literature on the Hubbard model~\cite{Tanizaki:2015rda}.

%%%%%%%%%%%%%%%%
\subsection{PIKfans for the one-site Hubbard model}
\label{sec:PIKfanExample}

As discussed in \Cref{sec:PIKfan}, the ripping of the manifolds shown in \Cref{fig:individual_zero_rip,fig:PIKfoldPlot} is expected from the general structure of the action. It cannot generally be avoided by any measure transport method that maps simple real distributions with a \textit{single} contour deformation to complex distributions with zeros. The PIKfan circumvents this obstruction by working with a mixture (i.e., a fan) of distributions and contour deformations. It thus enables us to take a fermionic theory of interest $p$ with a severe sign problem and faithfully map it to a distribution $p_0(\phi) = \sum_i w_i \rho_{i,0}(\phi)$ that has a similar sign problem. Importantly, this latter sign problem is chosen such that we can individually find real PIKfolds for each $\rho_{i,0}$. Here, we illustrate this with the toy example of the one-site Hubbard model.

To visualise the working principle of the PIKfan, we consider a general transformation that transports one Hubbard model distribution $p_0$ with the parameters $(U_0, \beta_0, \mu_0)$ into another Hubbard model distribution $p_1$ with the parameters $(U_1, \beta_1, \mu_1)$ by imposing transformations of the parameters $(U_t\,, \beta_t\,, \mu_t)$. One direct, illustrative way to ensure that $(E)$ and $(C)$ from \Cref{sec:PropertiesOfPIKfans} are fulfilled is to ensure that
\begin{align}
	\beta_t U_t = \textrm{(const.)} \,, \quad \beta_t\mu_t = \textrm{(const.)}\,.
	\label{eq:const_coupling_product_hubbard_model}
\end{align}
Then, for given parameters of the target distribution and an initial inverse temperature $\beta_0$, the path is uniquely defined by $\beta_t$. Here, we choose it to be
\begin{align}
	\beta_t = \beta_0 \left(\frac{\beta_1}{\beta_0} \right)^t \,,
\end{align}
as it provides a slowly varying kernel in the RG-time $t$. In particular, for this flow, the PIK $\dot\phi^{(w)}_t$ between $t=0$ and $t=1$ boils down to a temperature-dependent rescaling of the field values. This directly fulfils the conditions $(E)$ and $(C)$ from \Cref{sec:PropertiesOfPIKfans} as the map is regular everywhere.

In order to then compute observables with respect to the target distribution $p = p_1$, we follow the procedure described in \Cref{sec:RealSampling} and more explicitly in \labelcref{eq:SampleOtpM1}. Accordingly, we draw $N_s=5\times10^4$ samples from the initial real sampling pair $(\rho_{-1}, \mathcal{M})$ and transport them using the flow maps $\varphi^{(w)}_{i}$ on the respective ribs $i$ of the PIKfan. As explained around \Cref{fig:ComplexRealPIKfan}, these flow maps first transport the initial samples to the known complex Gau\ss ian decomposition at $t=0$. Subsequently, the samples are transported with the common PIK $\dot\phi^{(w)}_t$ to the final manifolds $\mathcal{M}_{i,1}$. Following \labelcref{eq:SampleOtpM1}, the final observables are then computed as a weighted sum of the observables on the individual manifolds $\mathcal{M}_{i,1}$ with the known positive mixture weights $w_i$ of the Gau\ss ian decomposition of $p_0$.

\begin{figure}[b]
	\centering
	\includegraphics[width=\linewidth]{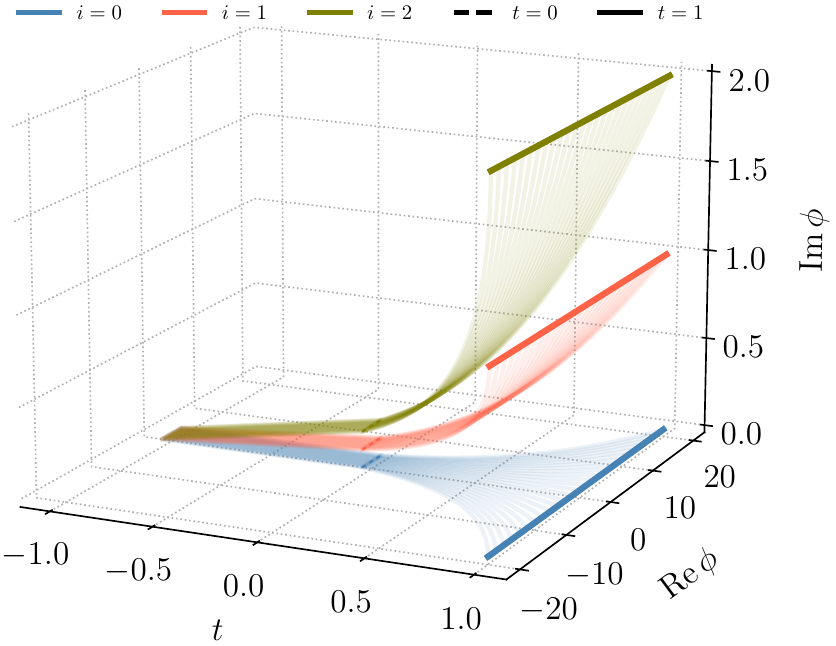}
	\caption{Evolution of the manifolds $\mathcal{M}_{i,t}$ from $t=-1$ to $t=1$ of the PIKfan for the flow of the distribution from \Cref{sec:PIKfanExample}.\hspace*{\fill}}
	\label{fig:hubbard_mixture_manifolds_3d}
\end{figure}

In order to illustrate this sampling along the PIKfan, we show the evolution of the manifolds in \Cref{fig:hubbard_mixture_manifolds_3d}. Here we choose the chemical potential $\mu_1=1.5$ for the target Hubbard model. This corresponds to the example from \Cref{fig:FirstRIP}, for which the na\"ive approach of using only a single PIK failed. For the initial distribution, we choose the inverse temperature $\beta_0=100$.

As all intermediate distributions are also Hubbard models, we can compute observables during the flow from $t=0$ to $t=1$ for different target chemical potentials $\mu_1$ (which define a corresponding initial distribution via \labelcref{eq:const_coupling_product_hubbard_model}).

\begin{figure}[t]
	\centering
	\includegraphics[width=0.49\textwidth]{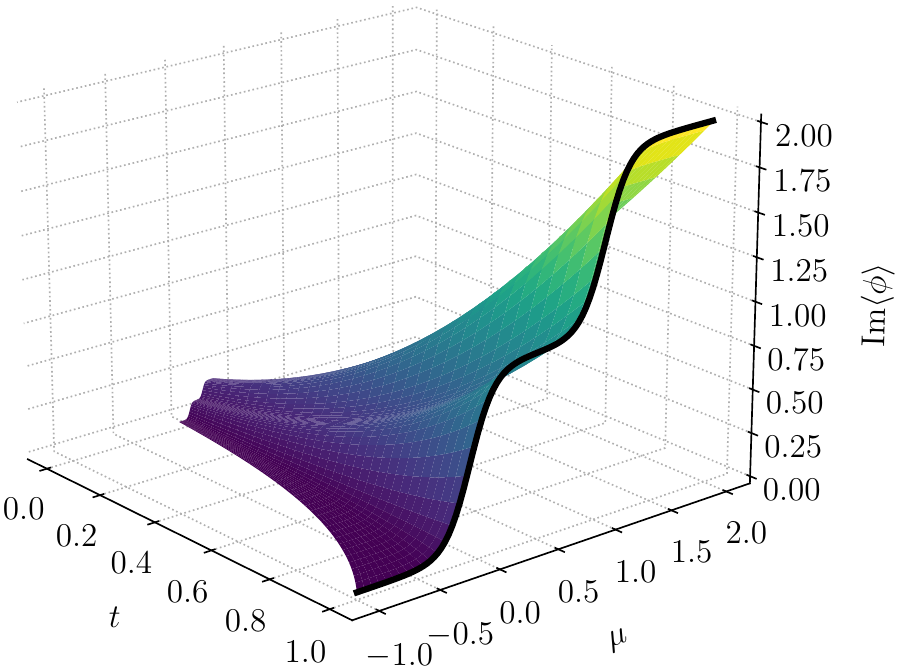}
	\captionof{figure}{Evolution of the observable $\mathrm{Im}\langle\phi\rangle$ (corresponding to the number density at $t=1$) for the flow of the distribution from \Cref{sec:PIKfanExample}. The flow transports along Hubbard models \labelcref{eq:HubbardModel} where $U_t,\,\beta_t,\,\mu_t$ change during the RG-time $t$. The black curve at $t=1$ shows the analytical result for the observable.\hspace*{\fill}}
	\label{fig:Im_phi_hubbard_pikfan}
\end{figure}

This procedure has been conducted for the same range of chemical potentials as in \Cref{fig:average_sign_hubbard}, and the results for the observable $\mathrm{Im}\langle \phi\rangle_{(p_t,\mathcal{M})}$ are shown in \Cref{fig:Im_phi_hubbard_pikfan}. We note that in this model at $t=1$ the number density is given by $n=-i\langle \phi\rangle/U=\mathrm{Im}\langle \phi\rangle / U$, see \labelcref{eq:density}. The analytically known result at $t=1$ from \labelcref{eq:HubbardA} is shown as a black line and is in agreement with the simulation.

Note again that as discussed in \Cref{sec:EquivalentSampling}, first and foremost, the PIKfan transforms a distribution $p$ with a severe sign problem into a distribution $p_0$ that is in the same equivalence class as $p$. Importantly, we choose a distribution $p_0$ whose structure renders its sign problem solvable, e.g.~a complex Gau\ss ian mixture. Then, provided the conditions from \Cref{sec:PropertiesOfPIKfans} are met, one can solve the sign problem of $p$.

Here, we illustrated this principle with the example of the one-site Hubbard model, which by itself decomposes into a Gau\ss ian mixture, see \Cref{app:analyticSol}.

We conclude this Section with a brief discussion of the Hubbard model beyond the simple one-site case. For the Hubbard model on a one-dimensional chain or a two-dimensional lattice, the present construction carries over, even though the Gau\ss ians obtained from the fermionic determinant come with positive and negative weights. Still, they can be pooled into positive distributions. While very interesting, the analysis of this case goes far beyond the current work, which concentrates on the construction of the PIKfan architecture, and will be presented elsewhere. 

%

%%%%%%%%%%%%%%%%%%%%%%%%%%%%%
\subsection{PIKfans beyond the Hubbard model}
\label{sec:phi4_mixture}

To exemplify the working principle of the PIKfan in a more involved example, we now consider a target distribution $p$ which cannot exactly be decomposed into a finite mixture of complex Gau\ss ians but comprises $\phi^4$ theories. 
This can be thought of as a simple toy model for higher-dimensional systems, in which higher-order terms in $\phi$ remain if the distribution is not decomposed completely into Gau\ss ians. 

To that end, we now consider the target distribution
\begin{align}
	p(\phi) = \frac{1}{\mathcal{N}}\sum_{i=1}^{N} w_i \hat g_i(\phi)\,,
	\label{eq:phi4_mixture}
\end{align}
where $\hat g_i(\phi)$ denotes the (unnormalised) distribution of a one-site $\phi^4$ theory with the quartic coupling $\lambda$, where the quadratic term is shifted into the complex plane as
\begin{align}
	\hat g_i(\phi) = \exp\left(-\frac{1}{2\sigma^2} (\phi - i m_i)^2 - \frac14 \lambda\, \phi^4\right) \,,
	\label{eq:phi4_distribution_component}
\end{align}
with $\lambda \in \mathbb{R}^+$, $\sigma, m_i \in \mathbb{R}$, and $\sum_i w_i=1$. At $\lambda=0$, the $w_i$ are the weights of the Gau\ss ian mixture $p^{(w)}$ in \labelcref{eq:WeightedSumDistribution1}. 
To investigate the transformation from the non-trivial distribution $p$ to a pure Gau\ss ian mixture $p_0$, we choose the path of the distribution where we make $\lambda$ time-dependent and tune it as
\begin{align}
	\lambda_t = \lambda t \,.
	\label{eq:lambda_evolution}
\end{align}
Then, we interpolate from a Gau\ss ian mixture at $t=0$ to a mixture of $\phi^4$ distributions at $t=1$. Having a similar situation as for the Hubbard model, we also take $N=3$ and the parameters $\{w_i\} = (0.5, 0.3, 0.2)\,,$ $ \{m_i\} = (0.0, 0.5, 1.0)\,,$ $\sigma = 1$, and $\lambda=1$. For this choice of parameters, the sign problem of the target distribution $p$ is still mild. The example should therefore be read as a proof of concept: it shows that the PIKfan applies directly to a broader class of distributions. Extending the approach to further (fermionic) theories and to regimes with a more severe sign problem (as done for the Hubbard model in \Cref{sec:PIKfanExample}) is part of future work.
\begin{figure}[t]
	\centering
	\begin{minipage}{0.49\textwidth}
		\centering
		\includegraphics[width=\linewidth]{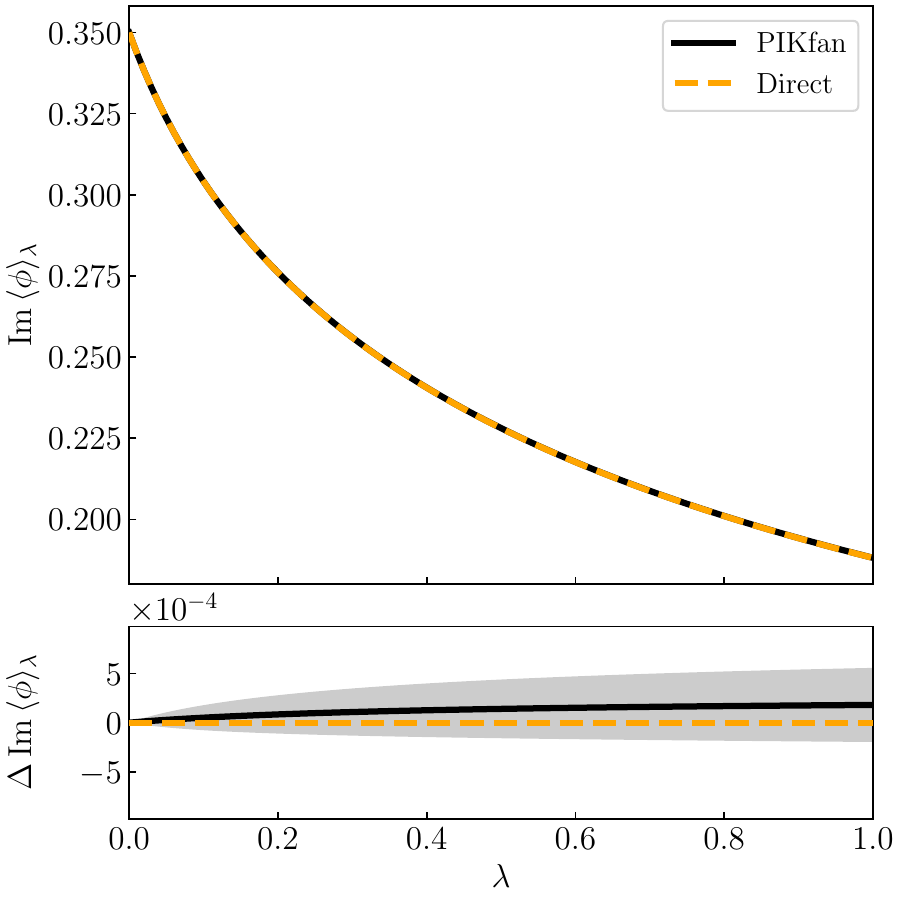}
		\captionof{figure}{Evolution of the observable $\mathrm{Im}\langle\phi\rangle$ for the flow of the distribution from \Cref{sec:phi4_mixture} computed with the PIKfan (black) and the direct integration (orange, dashed). The lower panel shows the relative deviation $\Delta\textrm{Im}\langle\phi\rangle_\lambda= (\textrm{Im}\langle\phi\rangle^\textrm{PIKfan}_\lambda-\textrm{Im}\langle\phi\rangle^\textrm{direct}_\lambda)/|\textrm{Im}\langle\phi\rangle^\textrm{direct}_\lambda|$, where the superscript `direct' indicates the result obtained from the numerical integration of the distribution on the real line. The grey band in the lower panel indicates the $1\,\sigma$ statistical error of the PIKfan mean.\hspace*{\fill}}
		\label{fig:phi4_mixture_Im_phi_t}
	\end{minipage}
\end{figure}	
Here, we can again easily sample from $p_0$ as it is a Gau\ss ian mixture. Then, solving for the PIKs between $t=0$ and $t=1$ induced by \labelcref{eq:phi4_mixture,eq:phi4_distribution_component,eq:lambda_evolution} allows us to compute observables of the distributions $p_t$ stochastically, although they form complex mixture distributions. In the present single-variable setting, the PIK map can be directly computed from \labelcref{eq:InverseCDFmap}. We compute $f_t$ by numerical quadrature and invert it with Newton's method on the samples of each rib. During the transport, Newton's method is initialised with the solution of the previous time step. This construction is specific to $d=0$, and general constructions are part of forthcoming work. In order to compute the observables, we follow the same approach as described in \Cref{sec:RealSampling} and conducted in \Cref{sec:PIKfanExample}.

The successful application of this procedure for $N_s = 2\times10^5$ samples is shown in \Cref{fig:phi4_mixture_Im_phi_t}. There we show the stochastically estimated $\mathrm{Im}\langle \phi\rangle_{\lambda}$ (black) together with the value computed by directly integrating the distribution (orange, dashed) on the real line. Here, the subscript $\lambda$ denotes the respective coupling value used in the distribution, which is a stand-in for the RG-time via \labelcref{eq:lambda_evolution}.

\begin{figure}
	\begin{minipage}{0.49\textwidth}
		\centering
		\includegraphics[width=\linewidth]{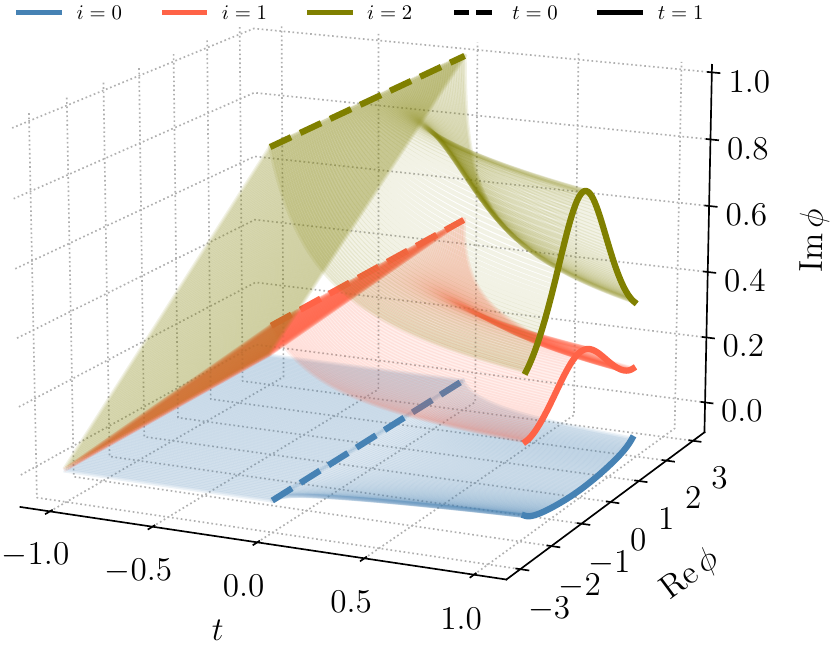}
		\captionof{figure}{Evolution of the manifolds $\mathcal{M}_{i,t}$ from $t=-1$ to $t=1$ of the PIKfan for the flow of the distribution from \Cref{sec:phi4_mixture}.\hspace*{\fill}}
		\label{fig:phi4_pikfan_manifolds_3d}
	\end{minipage}
\end{figure}

In order to illustrate the flow of the samples in the complex plane, we further show the time-evolution of the samples starting from the very first real Gau\ss ian distribution at $t=-1$. This directly illustrates the PIKfan in terms of the evolved manifolds and is shown in \Cref{fig:phi4_pikfan_manifolds_3d}. As for the one-site Hubbard model in \Cref{fig:hubbard_mixture_manifolds_3d}, the plot resembles the conceptual graphic introducing the PIKfan in \Cref{fig:ComplexRealPIKfan}. As discussed in \Cref{sec:PropertiesOfPIKfans}, the correctness of the observable computed in \Cref{fig:phi4_mixture_Im_phi_t} is ensured by $(E)$ and $(C)$ being fulfilled. Whereas $(E)$ is fulfilled by the smooth and non-singular evolution of the manifolds from $t=0$ to $t=1$, $(C)$ is met by the transport map remaining non-singular and uniquely defined between $\mathcal{M}_{i,1}$ and $\mathcal{M} = \mathbb{R}$. 

%

%%%%%%%%%%%%%%%%%%%%%
\section{Conclusion}
\label{sec:Conclusions}

In \cite{Ihssen:2026njd} the PIK architecture for solving sign problems for given target distributions $p(\phi)$ was put forward. Its pivotal property is the weight preservation of the PIK transport rooted in the Wegner equation, the functional RG equation for the distribution, which is nothing but a continuity equation. This architecture encompasses traditional approaches such as stochastic quantisation with complex Langevin equations and Lefschetz thimbles as specific projections. In contrast to these approaches, the PIK architecture comes with \textit{real} sampling on a flowing manifold ${\cal M}_t$ at each time step of the RG evolution, which is key to a constructive solution of sign problems. We have called this stack of manifolds with real sampling the \textit{real} PIKfold of the theory, see \Cref{sec:RealPIKfolds}. In particular, the PIK architecture does not require mapping out the complexified sampling manifold and its special points such as Lefschetz thimbles, zeros of the distributions or the global properties of the drift force. Instead, all the necessary information is \textit{computed} during the flow. In summary, this architecture allows the sampling of theories with sign problems at the computational cost of sampling a Gau\ss ian distribution and pushing the samples forward with the PIK. The computation of the PIK is either done with a given physics-informed basis or by simply using a stochastic solver for the Wegner equation with the samples themselves. The total cost is then that of simple (Gau\ss ian) sampling times the number of RG-time steps, together with the computation of the kernel itself, a price we pay gladly.

The PIK architecture described above requires the existence of a real PIKfold, which need not exist. This concerns in particular a large class of fermionic sign problems. In the present work we have extended the PIK architecture to sign problems without a real PIKfold, including fermionic ones. This generalisation completes the conceptual setup of the PIK architecture put forward in \cite{Ihssen:2026njd}. This task also required a classification of sign problems with equivalence classes of sampling pairs in \Cref{sec:EquivalentSampling}. Moreover, the requirement of real sampling led to the introduction of the PIKfan in \Cref{sec:RealSampling,sec:PropertiesOfPIKfans}. In short, a PIKfan consists of a collection of real sampling pairs that allow one to sample a weighted sum of (simple) complex distributions. This sum of simple distributions is then mapped to the target distribution with a PIK. This construction reduces the task of solving sign problems for a given target distribution to that of finding a sum of simple complex distributions with positive weights, each of which admits a real PIKfold, that lies in the same equivalence class as the target. Such decompositions with positive weights exist for all theories with a statistical interpretation, as follows from the spectral representation \labelcref{eq:SpectralZ} of the generating function, see \Cref{sec:RealSampling}. For the one-site Hubbard model, the weights are the Boltzmann factors of the Fock states in \labelcref{eq:HubbardA}, see \labelcref{eq:WeightsHubbard}.

The working principle of the PIKfan was illustrated with the one-site Hubbard model in \Cref{fig:hubbard_mixture_manifolds_3d} and with a $\phi^4$-deformed extension in \Cref{fig:phi4_pikfan_manifolds_3d}. The PIKfan allows us to sample the one-site Hubbard model for general chemical potential with importance sampling of a real Gau\ss ian distribution. In this example, the PIK between the weighted sum and the target is a temperature-dependent rescaling of the field, see \Cref{sec:PIKfanExample}. PIKs beyond rescalings are required for the $\phi^4$-deformed model in \Cref{sec:phi4_mixture}. All examples in the present work have a single variable, $|\mathcal{D}|=1$ ($d=0$), for which the zeros of the distribution are isolated points. For $|\mathcal{D}|\geq2$, the zero sets are extended, see \Cref{sec:zeros_and_transport}, and their intersections with the sampling manifolds generically persist over finite intervals of the RG-time.

Altogether, the PIKfan architecture set up in the present work constitutes a very promising approach for sampling fermionic theories with sign problems. First applications to the Hubbard model in one and two dimensions with small and moderate lattice volumes are under way, and the results will be presented elsewhere.

The application of the PIKfan architecture to fermionic theories on larger lattices requires the resolution of some remaining challenges: the efficient construction of positive decompositions, whose number of terms grows exponentially with the volume in the exact ansatz of \Cref{app:analyticSol}, the control of extended zero sets for $|\mathcal{D}|\geq2$, and, for a stochastic solver, the evaluation of the kernel away from the samples, as required for the transport of the zeros of $p_0$, see \labelcref{eq:field_transport_zero_transport_app,eq:zeros_along_transport}. In particular, the PIKfan has to be constructed rather than searched for, see \Cref{sec:PropertiesOfPIKfans,sec:FindingPIKfans} for more details. We have set up such a constructive procedure with emergent PIKfans, and we hope to report on its application to higher-dimensional models soon. A further interesting step is the application of the PIKfan architecture to gauge-fermion systems including QCD and QCD-like theories. This is work in progress and also uses the emergent PIKfan architecture. Finally, the task of the PIKfan search as well as the choice of the path in action space is tailor-made for optimisation with machine learning. Moreover, the close relation of the PIK approach to diffusion models and normalising flows, see e.g.~\cite{Song:2020hus,Fukushima:2024oij,Aarts:2025lpi,Ihssen:2025hyl,Sexty:2026aym,Grinwald:2026phx}, suggests a combined application, which is also the subject of work in progress.

%%%%%%%%%%%%%%%%%%%%%%%%%
\section*{Acknowledgements}

We thank Manfred Salmhofer, Christian Schmidt, Yang-yang Tan, Finn Temmen, Lingxiao Wang and Kai Zhou and the members of the DM-QFT collaboration for discussions. This work is funded by the Deutsche Forschungsgemeinschaft (DFG, German Research Foundation) under Germany’s Excellence Strategy EXC 2181/1 -- 390900948 (the Heidelberg STRUCTURES Excellence Cluster), the Collaborative Research Centre SFB 1225 -- 273811115 (ISOQUANT), the Collaborative Research Centre SFB 1238 -- 277146847 (Control and Dynamics of Quantum Materials) and the Walter Benjamin Programme with project number 574031240. This work is also supported in part by Japan Society for the Promotion of Science (JSPS) KAKENHI Grant Nos.\ 22H05118, 25K24464, 26K00698. The authors used the AI assistant Claude (Anthropic) for proofreading, consistency checks of equations and notation, the discussion of examples, and internal reviews of the draft, and ChatGPT (OpenAI) for suggestions on clarity. All scientific content and conclusions are the authors' own, and the authors take full responsibility for the manuscript.

	\begingroup
		\appendix
\allowdisplaybreaks

%%%%%%%%%%%%%%%%%%%%%%%%%%%%%
\section{Detailed example of a ripping manifold}
\label{sec:NoRip}

In \Cref{sec:RippingExample} we introduced a simple example with a ripping sampling manifold, recall \labelcref{eq:distribution_D_problem} and \Cref{fig:individual_zero_rip}. Here, the existence condition $(E)$ from \Cref{sec:PropertiesOfPIKfans} was violated: the ripped $\mathcal{M}_t$ no longer constitutes a sampling manifold for the computation of observables via \labelcref{eq:SampleOtpM1}. In this Appendix, we use the analytical tractability of this example to discuss the ripping of the manifold in more detail and to illustrate the underlying mechanism.

Recall from \Cref{sec:RippingExample} that the distribution $p_t$ of this example from \labelcref{eq:distribution_D_problem} has a moving zero $z_t = ia/t$ for $t>0$. At this zero, the kernel $\dot\phi_t$ from \labelcref{eq:D_problem_kernel} has a pole. When the manifold $\mathcal{M}_t$ sweeps over this pole, the manifold rips, as seen in \Cref{fig:individual_zero_rip}.

We begin our detailed discussion by visualising the error caused by the ripping manifold for the computation of observables. To that end, we consider
\begin{align}
	\langle \mathcal{O}\rangle^{\phantom{1}}_{\textrm{flow}, t} := \frac{1}{N_s} \sum_{n=1}^{N_s} \mathcal{O}(\, \phi_t(\phi_n)\, ) \,,
	\label{eq:observable_flow_map_app}
\end{align}
with the flow map $\phi_t$ from \labelcref{eq:phiMap} and $\phi_n\sim p_0$ irrespective of whether $\mathcal{M}_t = \phi_t(\mathbb{R})$ rips. Hence, \labelcref{eq:observable_flow_map_app} relaxes the requirement of $(E)$ that we demand for \labelcref{eq:SampleOtpM1}. In \Cref{fig:E_problem_observable}, we compare the real (blue) and imaginary (red) parts of $\langle \phi^2\rangle^{\phantom{1}}_{\textrm{flow}, t}$, estimated with $N_s=5\times 10^4$ samples, with the exact $\langle\phi^2\rangle_{(p_t,\mathbb{R})}$ (black) obtained by direct integration. Both agree perfectly at early times, but deviate once $\mathcal{M}_t$ hits the zero $z_t$ at $t_{\textrm{rip}}/a =0.7651\ldots$.
\begin{figure}[t]
	\centering
	\includegraphics[height=\linewidth]{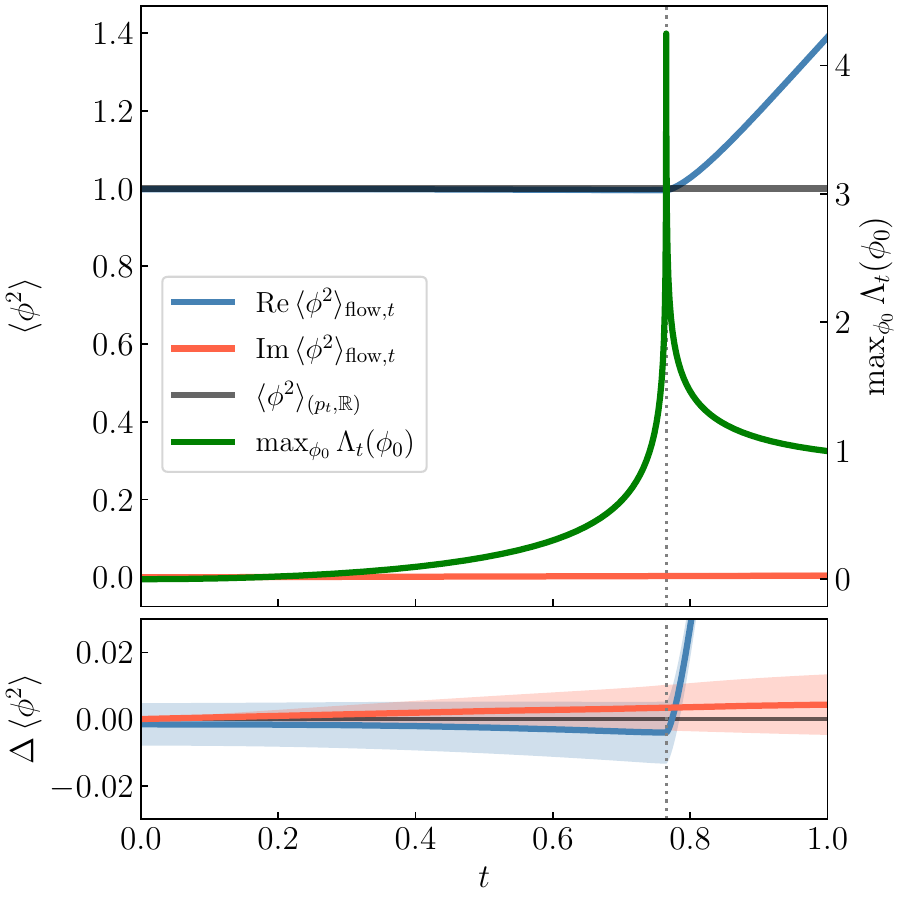}
	\caption{Real (blue) and imaginary (red) parts of $\langle \phi^2 \rangle$ for the distribution $p_t$ from \labelcref{eq:distribution_D_problem} as a function of the RG-time $t$, computed with the PIK \labelcref{eq:D_problem_kernel} and compared with the exact result (black). In addition, the regularity indicator $\max_{\phi_0}\Lambda_t(\phi_0)$ from \labelcref{eq:regularity_indicatorApp} (green) is shown. The lower panel shows the relative deviation $\Delta\langle\phi^2\rangle_t = \left( \langle\phi^2\rangle^{\textrm{PIK}} - \langle\phi^2\rangle^{\textrm{direct}} \right) / |\langle\phi^2\rangle^{\textrm{direct}}|$, where the superscript `direct' indicates the result obtained from the numerical integration of the distribution on the real line. The band in the lower panel indicates the $1\,\sigma$ statistical error of the PIK mean.\hspace*{\fill}}
	\label{fig:E_problem_observable}
\end{figure}

The failure of $(E)$ can also be monitored directly. As introduced in \labelcref{eq:regularity_indicator}, a good scalar indicator for the ripping is the maximum over the samples, $\max_{\phi_0\sim p_0}\Lambda_t(\phi_0)$, of the modulus of the log-Jacobian accumulated during the flow. We recall \labelcref{eq:regularity_indicator} here for the sake of convenience, 
\begin{align}
	\Lambda_t(\phi_0) := \left| \int_0^t ds \, \frac{\partial \dot\phi_{s,j}(\phi_{s})}{\partial\phi_{j}} \right|\,.
	\label{eq:regularity_indicatorApp}
\end{align}
As the transport is weight-preserving, $\Lambda_t$ directly tests whether the manifold rips in regions that contribute significantly to the observable. Indeed, \Cref{fig:E_problem_observable} shows that the maximum of $\Lambda_t$ (green) peaks when the computation of the observable fails and the manifold rips.

In zero dimensions, the role of the zero can be further illustrated with the cumulative distribution function (CDF) of $p_t$,
\begin{align}
	f_t(\phi) := \int_{-\infty}^{\phi} d\phi' \, p_t(\phi') \,, \qquad \frac{\partial f_t(\phi)}{\partial \phi} = p_t(\phi) \,.
	\label{eq:E_problem_cdf}
\end{align}
Since the flow map is weight-preserving, we have
\begin{align}
	f_0(\phi) = f_t(\,\phi_t(\phi)\,) \,.
\end{align}
Thus, we can compute the flow map also via the inverse CDF $f_t^{-1}$ as
\begin{align}
	\phi_t(\phi) &=  f_t^{-1}(\,f_0(\phi)\,) \,.
	\label{eq:E_problem_flow_map}
\end{align}
However, by \labelcref{eq:E_problem_cdf}, the zero $z_t$ of the distribution is a critical point of the CDF $f_t$. Thus, in the vicinity of $z_t$, we can express the CDF as
\begin{align}
	f_t(\phi) = f_t(z_t) + \frac{1}{2} \frac{\partial p_t(z_t)}{\partial\phi} (\phi - z_t)^2
	+ \mathcal{O}(\, (\phi - z_t)^3\, ) \,.
\end{align}
Hence, $f_t$ is not injective in the vicinity of $z_t$, and its inverse $f_t^{-1}$, which enters the flow map \labelcref{eq:E_problem_flow_map}, has a square-root branch point at $f_t(z_t)$. This is shown explicitly in \Cref{fig:E_problem_inverse_cdf}: for a CDF value $s:= f_t(z_t) + \epsilon$ close to the critical value, the curves $\mathrm{Re}\, f_t(\phi) = \mathrm{Re} \, s$ and $\mathrm{Im}\, f_t(\phi) = \mathrm{Im}\, s$ intersect at two points, which shows that $f_t^{-1}$ is double-valued.
\begin{figure}[t]
	\centering
	\includegraphics[width=\linewidth]{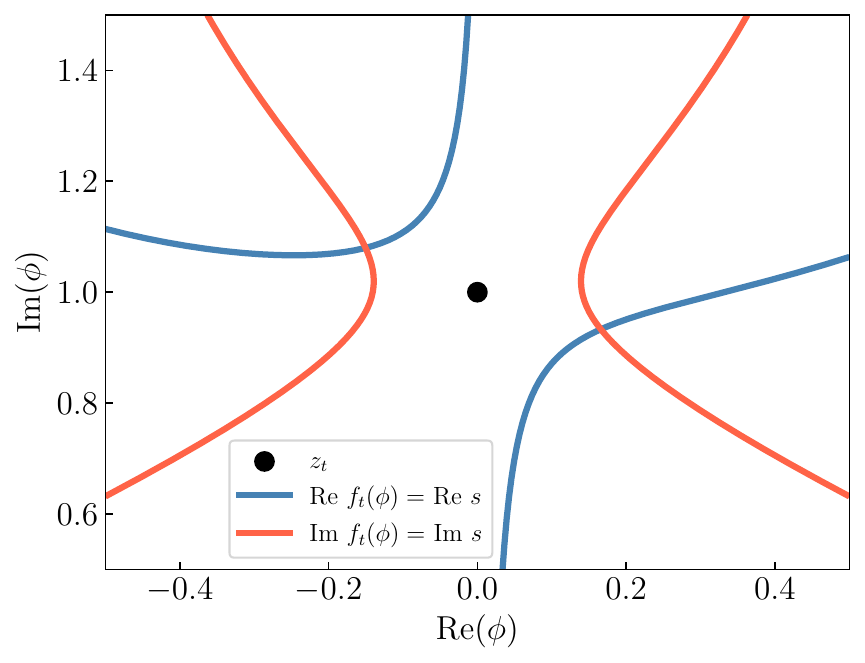}
	\caption{Visualisation of the multivaluedness of the inverse CDF $f_t^{-1}$ in the vicinity of $f_t$'s critical point $z_t$, where $p_t(z_t) = 0$. Given $s := f_t(z_t) + \epsilon$ with $|\epsilon| \ll 1$, the blue/red line shows the match of the real/imaginary part of $f_t(\phi)$ in the vicinity of $z_t$. Intersection points of the blue and red lines determine the two values of $f_t^{-1}(s)$.\hspace*{\fill}}
	\label{fig:E_problem_inverse_cdf}
\end{figure}
%

%%%%%%%%%%%%%%%%%%%%%%%%%%%%
\section{Tracking the requirements of Cauchy's theorem with PIKs}
\label{sec:PIKfanCauchy}

In \Cref{sec:RippingExample,sec:NoRip}, the manifold rips and condition $(E)$ is violated in a tractable manner. As no singularities are present between $\mathcal{M}_t$ and $\mathcal{M}$, these examples isolate $(E)$ but do not probe $(C)$. In this Appendix, we consider the reverse situation: the intermediate and final manifolds $\mathcal{M}_{i,t}$ exist, and thus $(E)$ is fulfilled, but $(C)$ is violated.

As a concrete example, we choose the one-site Hubbard model from \Cref{sec:OneSiteHubbard} as the target distribution $p=p_1$. It is a sum of three complex Gau\ss ians with the weights $w_i\in \mathbb{R}^{+}$ of \labelcref{eq:WeightsHubbard}, $\sum_i w_i=1$, a shared real variance $\sigma^2=U/\beta$, and equally spaced means $m_i = (i-1) \Delta m$ with $\Delta m = U$, see \Cref{sec:OneSiteHubbard,sec:variable_transformations_hubbard},
\begin{align}
	p(\phi) = \frac{1}{\sqrt{2\pi\sigma^2}}\sum_{i=1}^{3} w_i \exp[-\frac{1}{2\sigma^2}\left( \phi- i m_i \right)^2]\,.
	\label{eq:C_problem_distribution}
\end{align}

In order to test violations of $(C)$, we introduce a path along which the weights $w_i$ are kept fixed, while the variance and the spacing of the means change according to
\begin{align}
	\sigma_t = \sigma \delta_t \,, \quad \Delta m_t = \Delta m \delta_t^\kappa\,,
	\label{eq:C_problem_path_dist_parameters}
\end{align}
with 
\begin{align}
	\delta_t = \delta^{1-t}\,,
	\label{eq:C_problem_path_delta}
\end{align}
so that $p_1$ is the chosen Hubbard model. Both $\delta$ and $\kappa$ are hyperparameters of the path. We use $U=1\,,\beta=10$ and $\mu=1.2$, which corresponds to a severe sign problem, see \Cref{fig:average_sign_hubbard}. For the path parameters, we choose $\delta=0.8$ and take the two choices $\kappa = 1.0$ and $\kappa =0.9$.

This path is motivated by the zeros of the distribution. In the Hubbard--Stratonovich transformation used here, the zeros of the one-site Hubbard model have multiplicity two. Along the path, the zeros of $p_t$ remain double zeros for all $t$ if $\kappa = 1$. For $\kappa \neq 1$, they are simple zeros for $t<1$, and pairs of them have to merge into double zeros at $t=1$. As shown in \Cref{sec:zeros_and_transport}, a regular flow maps zeros onto zeros, and an injective one cannot merge them. We therefore expect $(C)$ to hold for $\kappa = 1.0$ but to fail for $\kappa = 0.9$.

\begin{figure}[t]
	\centering
	\includegraphics[width=0.49\textwidth]{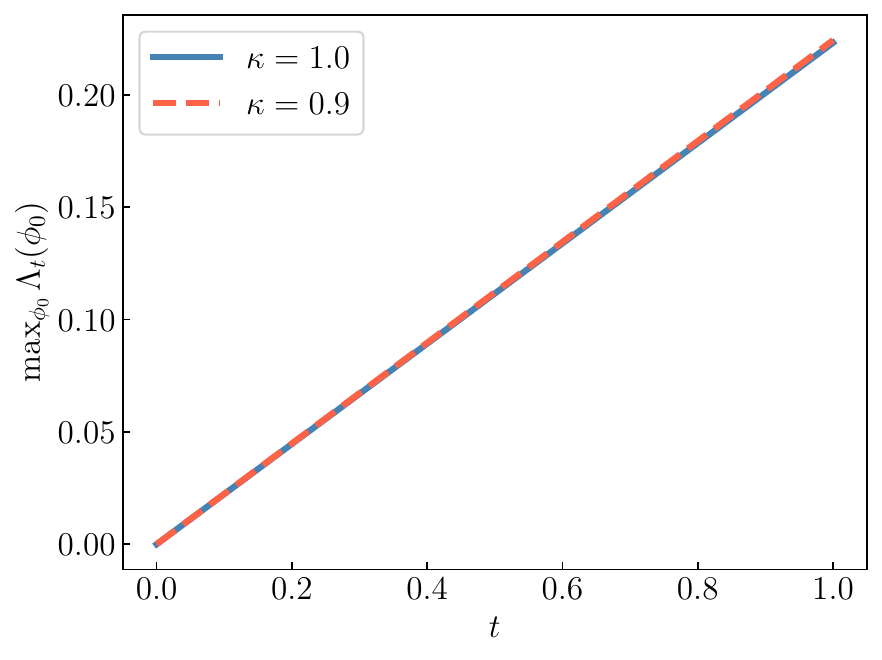}
	\caption{Regularity indicator $\max_{\phi_0} \,\Lambda_t(\phi_0)$ from \labelcref{eq:regularity_indicatorApp} along the RG-time $t$ for the two choices of $\kappa$ from \labelcref{eq:C_problem_path_dist_parameters}, showing that condition $(E)$ is fulfilled for the distribution path given by \labelcref{eq:C_problem_distribution,eq:C_problem_path_dist_parameters,eq:C_problem_path_delta}.\hspace*{\fill}}
	\label{fig:C_problem_D_problem}
\end{figure}

We first confirm that $(E)$ is fulfilled for both values of $\kappa$. To this end, we monitor $\Lambda_t$ from \labelcref{eq:regularity_indicatorApp}, as in \Cref{sec:NoRip}. \Cref{fig:C_problem_D_problem} shows that $\Lambda_t$ remains finite for both values of $\kappa$; the two curves in fact overlap. Hence, $(E)$ is fulfilled, and the manifolds $\mathcal{M}_{i,t}$ are smooth deformations of $\mathcal{M}$ as $p_t$ is deformed from $p_0$ to $p_1$.

For $\kappa = 1.0$, the flow map $\phi_t$ is injective, also in the regions that are not swept over when pushing the samples from $\mathcal{M}_{i,0}$ to $\mathcal{M}_{i,1}$, and each zero of the distribution is mapped onto exactly one zero, see \Cref{fig:C_problem_manifolds_kappa_10}. There, as in \Cref{fig:C_problem_manifolds_kappa_09}, the zero sets are obtained by transporting the known zeros of $p_0$ with the computed kernel $\dot\phi^{(w)}_t$, see \labelcref{eq:field_transport_zero_transport_app}; no search for zeros is involved. Hence, the $\rho_{i,1}$ encounter no singularity when Cauchy's theorem is applied, and the observable in \Cref{fig:C_problem_obs_kappa_10} agrees with the exact values along the entire flow.

For $\kappa = 0.9$, in contrast, the observable computed with the PIKfan deviates from the exact values beyond the Monte Carlo error, although $(E)$ is fulfilled, see \Cref{fig:C_problem_obs_kappa_09}. The reason is seen in \Cref{fig:C_problem_manifolds_kappa_09}: while the sampling manifolds evolve as smoothly as for $\kappa = 1.0$, pairs of simple zeros in the region between $\mathcal{M}_{3,1}$ and $\mathcal{M}=\mathbb{R}$ merge into double zeros at $t=1$. Hence, the flow map, which maps two distinct zeros onto one, cannot be injective there. As discussed in \Cref{sec:zeros_and_transport}, $\rho_{3,1}$ then picks up a branch point, which violates Cauchy's theorem $(C)$.
\begin{figure*}[t]
	\centering
	\begin{subfigure}{0.49\textwidth}
		\centering
		\includegraphics[height=0.6\textwidth]{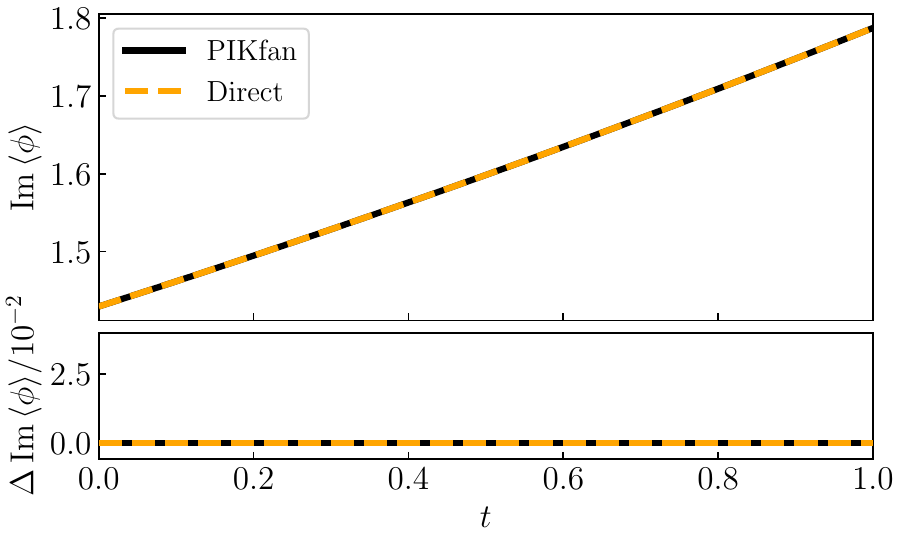}
		\caption{Observable for $\kappa = 1.0$.}
		\label{fig:C_problem_obs_kappa_10}
	\end{subfigure}%
	\hfill%
	\begin{subfigure}{0.49\textwidth}
		\centering
		\includegraphics[height=0.6\textwidth]{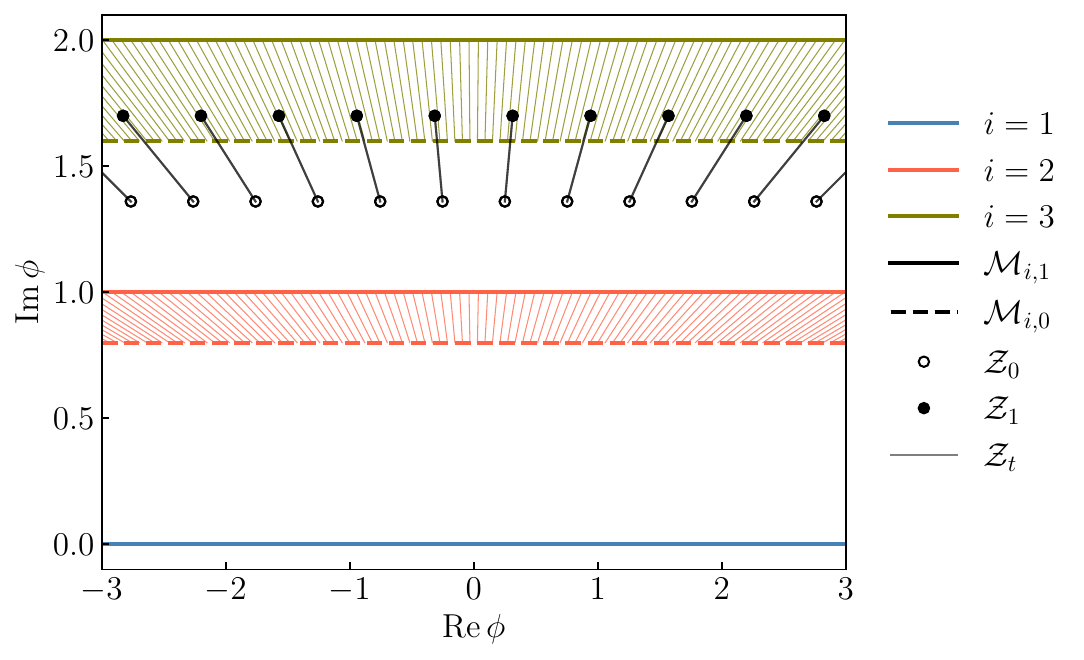}
		\caption{Manifolds and zero sets for $\kappa = 1.0$.}
		\label{fig:C_problem_manifolds_kappa_10}
	\end{subfigure}
	\\[2ex]
	\begin{subfigure}{0.49\textwidth}
		\centering
		\includegraphics[height=0.6\textwidth]{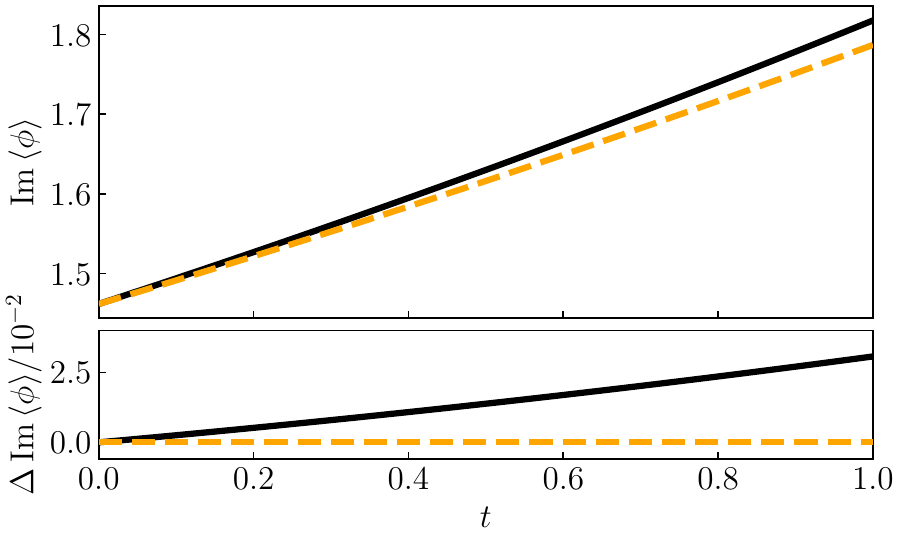}
		\caption{Observable for $\kappa = 0.9$.}
		\label{fig:C_problem_obs_kappa_09}
	\end{subfigure}%
	\hfill%
	\begin{subfigure}{0.49\textwidth}
		\centering
		\includegraphics[height=0.6\textwidth]{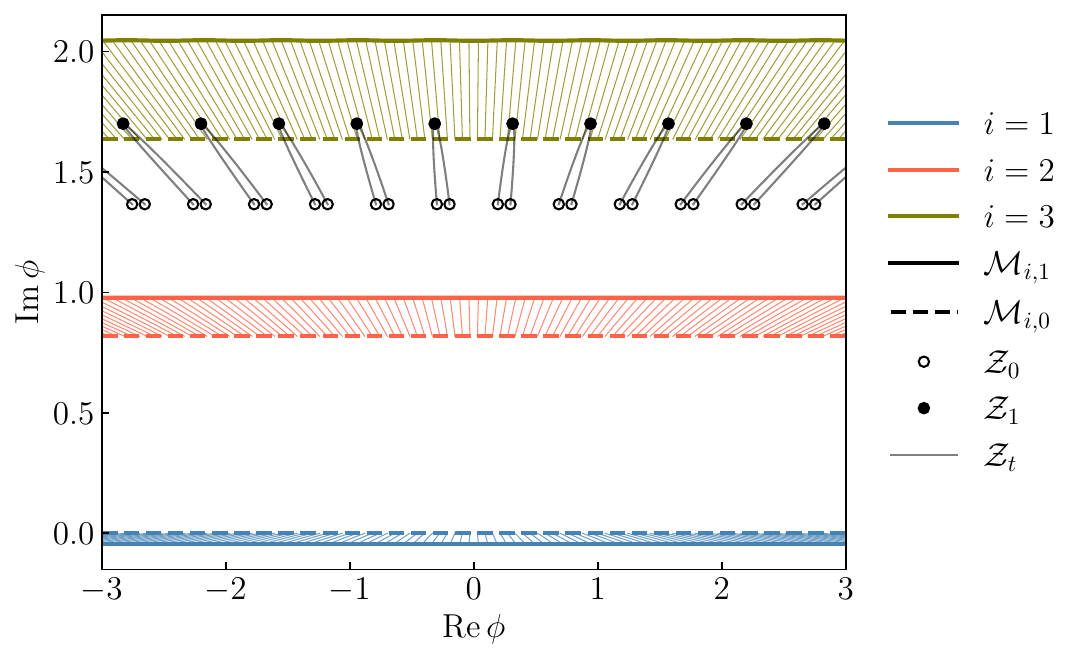}
		\caption{Manifolds and zero sets for $\kappa = 0.9$.}
		\label{fig:C_problem_manifolds_kappa_09}
	\end{subfigure}
	\caption{Results for the path given by \labelcref{eq:C_problem_distribution,eq:C_problem_path_dist_parameters,eq:C_problem_path_delta} with $\beta = 10\,,U=1\,,\mu=1.2$ and $\delta=0.8$, for $\kappa = 1.0$ (top) and $\kappa=0.9$ (bottom). The left column shows the observable $\mathrm{Im}\langle \phi\rangle$ during the flow computed with the PIKfan (black) and by direct integration (orange, dashed). On the left, each lower panel shows the relative deviation $\Delta\,\mathrm{Im}\langle\phi\rangle$ of the PIKfan result from the direct integration, where the normalisation was taken to be the modulus of the direct result. The right column shows the evolution of the manifolds $\mathcal{M}_{i,t}$ (coloured) and the zero set $\mathcal{Z}_t$ of $p_t$ (black) under the transport induced by $\dot\phi_t^{(w)}$. \hspace*{\fill}}

\end{figure*}
%
%%%%%%%%%%%%%%%%%%%%%%%%%%%%%
\section{Analytic solution of the one-site Hubbard model}
\label{app:analyticSol}
\begin{figure*}
	\begin{subfigure}{.49\linewidth}
		\includegraphics[width=\linewidth]{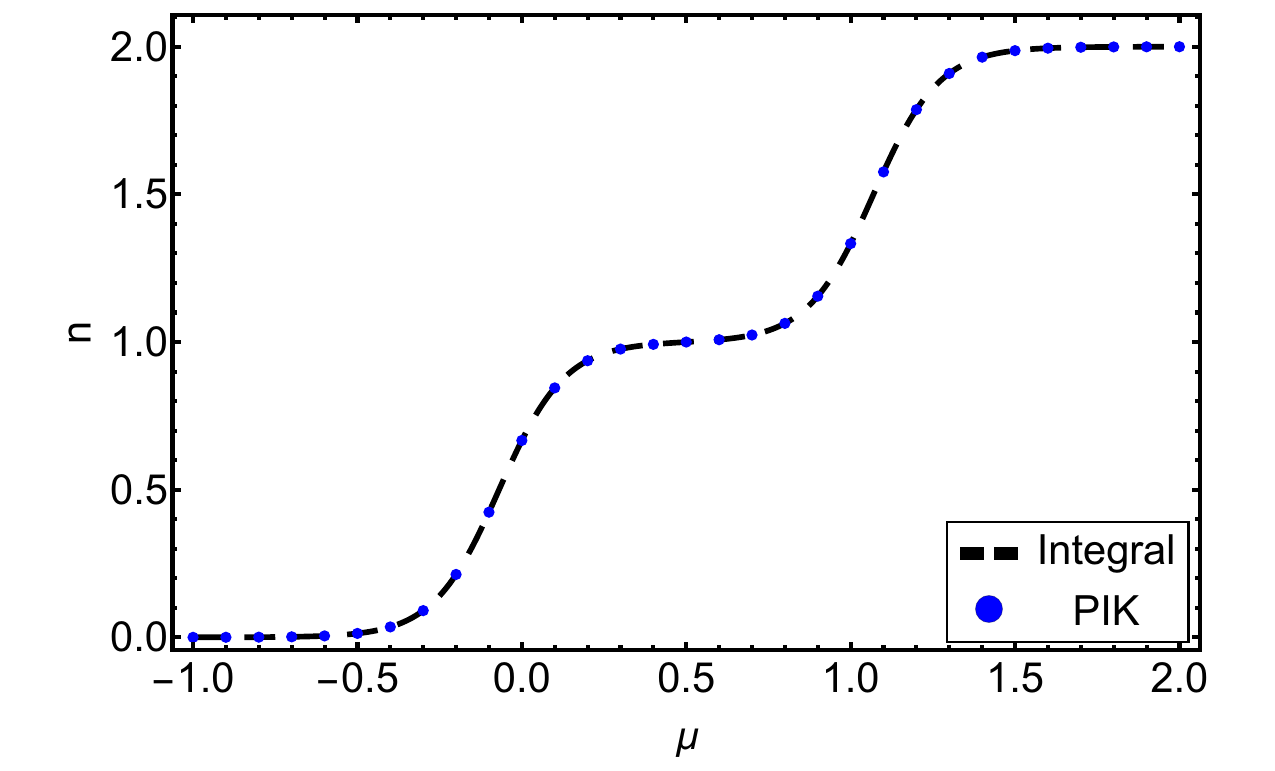}
                \captionsetup{width=.85\linewidth,justification=centering}
		\subcaption{Sampled result from the PIKfan in comparison to the exact solution.\hspace*{\fill}}
	\end{subfigure}
	\begin{subfigure}{.49\linewidth}
		\includegraphics[width=\linewidth]{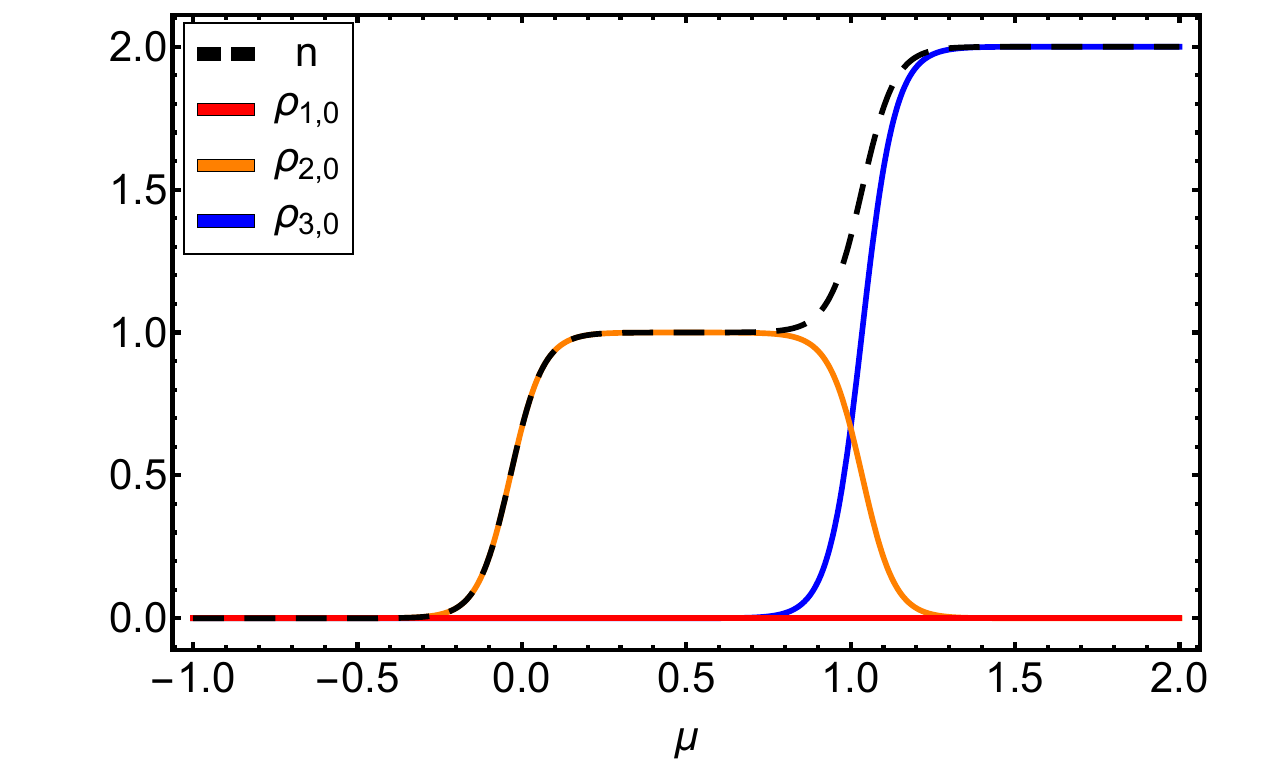}
                \captionsetup{width=.85\linewidth,justification=centering}
		\subcaption{Contributions from the distinct Gau\ss ians to the PIKfan \labelcref{eq:SplitRho}.\hspace*{\fill}}
		\label{fig:densityPlot2}
	\end{subfigure}
	\centering
	\caption{Density \labelcref{eq:density} as a function of chemical potential $\mu$ as computed from the solution of the PIKfan \labelcref{eq:SumOperator} for couplings $U=1$ and $\beta=10$. The region $-U/2 \lesssim \mu \lesssim 3U/2$ is affected by a particularly severe sign problem.\hspace*{\fill}}
	\label{fig:densityPlot}
\end{figure*}

In this Appendix, we discuss the analytic structure of the solution to the one-site Hubbard model directly from the path integral and show how this translates to the splitting of the PIKfan.

Considering the shape of the distribution from \labelcref{eq:HubbardZ}, we find three distinct Gau\ss ian contributions
\begin{subequations}
	\label{eq:SumZ} 
	\begin{align} 
		Z=Z_1+Z_2+Z_3\,,
		\label{eq:ZHubbard}
	\end{align}
	with 
	\begin{align}\nonumber 
		Z_1 &= \frac{1}{\mathcal{N}}  \int d\phi\, e^{-\frac{\beta}{2U}\phi^2}\,,\\[1ex]\nonumber 
		Z_2 &= \frac{2}{\mathcal{N}}  \int d\phi\, e^{\beta \left(\mu + \frac{U}{2} + i \phi \right) -\frac{\beta}{2U}\phi^2}\,,\\[1ex]
		Z_3 &= \frac{1}{\mathcal{N}} \int d\phi\, e^{2 \beta \left(\mu + \frac{U}{2} + i \phi \right) -\frac{\beta}{2U}\phi^2}\,.
		\label{eq:Split}
	\end{align}
\end{subequations}
The integrands in \labelcref{eq:Split} follow from expanding $e^{-\hat S(\phi)}$ in \labelcref{eq:HubbardModel}, and $\mathcal{N}$ is the normalisation \labelcref{eq:NormalisationHubbard}. The first integral $Z_1$ is simply a real constant.
The next one, $Z_2$, looks complex, but noting that
\begin{align}
\frac{\beta}{2U}\phi^2 - \beta \left(\mu + \frac{U}{2} + i \phi \right) = \frac{\beta}{2U}(\phi - iU)^2 - \beta\mu \,,
\end{align} 
we can make a variable shift as $\phi  \to \phi+ iU$ to find that the Gau\ss ian integral is proportional to $2e^{\beta\mu}$, which is nothing but the second term in \labelcref{eq:HubbardA}, up to the common factor $\sqrt{2\pi U/\beta}\,\big/\mathcal{N}$ shared by all $Z_i$.  We can shift $\phi\to \phi+2iU$ for $Z_3$ to reproduce the third term in \labelcref{eq:HubbardA}, up to the same factor.
 
 The shifted Gau\ss ians already correspond to the structure we require to apply the \textit{PIKfan} with $t\in[-1,0]$: we skip the second step with $t\in[0,1]$ as we recover the Hubbard model at $t=0$. This is particularly interesting in regions with a severe sign problem, $-U/2 \lesssim \mu \lesssim 3U/2$, where the physical action is no longer connected to real sampling on the real axis $\mathcal{M}_0=\mathbb{R}$.
For the one-site Hubbard model \labelcref{eq:HubbardZ}, the action can be disentangled and rewritten as a sum of three Gau\ss ian distributions \labelcref{eq:WeightedSumDistribution} with
\begin{align}\nonumber 
	\rho_{1,t} (\phi)= &\,\frac{1} {\mathcal{N}_{1,t}} e^{-\frac{\beta}{2U}\phi^2}  \,,  \\[1ex]\nonumber 
	\rho_{2,t} (\phi)= &\,\frac{1} {\mathcal{N}_{2,t}}  e^{(1+t) \beta \left(\mu + \frac{U}{2} + i \phi \right) -\frac{\beta}{2U}\phi^2} 
	\,,\\[1ex]
	\rho_{3,t} (\phi)= &  \, \frac{1} {\mathcal{N}_{3,t}}  e^{2(1+t) \beta \left(\mu + \frac{U}{2} + i \phi \right) -\frac{\beta}{2U}\phi^2}    \,,
\label{eq:SplitRho}
\end{align}
with $t\in [-1,0]$. The weights $w_i$ of the distribution $p^{(w)}$ in \labelcref{eq:WeightedSumDistribution1} are given by  
\begin{align} 
	w_{i} = \frac{a_i\,\mathcal{N}_{i,0} }{ \sum_j a_j\,\mathcal{N}_{j,0} }\,\quad \textrm{with}  \quad i,j=1,2,3\,.
	\label{eq:WeightsHubbard}
\end{align}
The multiplicities of the Gau\ss ians are given by $a_1=a_3=1$, $a_2=2$  and the
$\mathcal{N}_{i,t}$ are the corresponding normalisations. This corresponds to the first step in \Cref{fig:ComplexRealPIKfan}.

At $t=-1$, all three distributions correspond to the same single real Gau\ss ian, whereas at $t=0$, we already recover the physical action of the one-site Hubbard model. Note that this exact ansatz is also possible for larger lattices in the Hubbard model. However, the number of required Gau\ss ians grows exponentially with the lattice volume, like the number of terms in the fermion determinant.

Each of the distributions in \labelcref{eq:SplitRho} corresponds to a Gau\ss ian and allows for the analytical computation of a sign-problem-free contour using a real PIKfold. The calculation of a PIKfold for a Gau\ss ian with a linear current term was performed explicitly in Supplement 3.1 of \cite{Ihssen:2026njd}, and results in a linear shift of the contour into the complex plane, which we already observed below \labelcref{eq:Split}.

The solution to the \textit{PIKfan} is given by
\begin{align}
	\dot \phi_{1,t} &=0 &\Rightarrow \quad \phi_{1,t}(\phi) &= \phi \,,\notag\\[1ex]
	\dot \phi_{2,t} &=i U &\Rightarrow \quad \phi_{2,t}(\phi) &= \phi + i U(1+t) \,,\notag\\[1ex]
	\dot \phi_{3,t} &=2i U &\Rightarrow \quad  \phi_{3,t}(\phi)& = \phi + 2 i U (1+t)\,,
\end{align}
where $\phi = \phi _{-1} \in \mathbb R$ is the variable sampled at $t=-1$ with 
\begin{align}
	\phi \sim  e^{-\frac{\beta}{2U}\phi^2} \,.
\label{eq:GaussSample}
\end{align}
By weighting the total distribution with the adjusted normalisations, we can now sample general holomorphic operators $\mathcal{O}(\phi)$ with
\begin{align}
	\langle \mathcal{O} \rangle_{(p,\mathcal{M})} \simeq\frac{1}{N_{s}}\sum_n \sum_i  w_i\, \mathcal{O}(\phi_{i,0}(\phi_n)) \,,
\label{eq:SumOperator}
\end{align}
where $N_{s}$ is the total number of Gau\ss ian samples $\phi_n$ and $i=1,2,3$. As an illustration of our result, we consider the number density
\begin{align}
n = \frac{1}{\beta}\frac{\partial \ln \mathcal{N}}{\partial\mu} =  -i \langle \phi \rangle / U \,.
\label{eq:density}
\end{align}
The $\mu$-dependence of the density at $\beta= 10$ and $U=1$ is displayed in \Cref{fig:densityPlot}; the \textit{PIKfan} result is obtained with a sample size $N_{s}=10^4$. We also show the solution of a direct evaluation of the integral as a comparison.
In \Cref{fig:densityPlot2} we show the different contributions of the single distributions $w_i \rho_i$ to the total density.

%%%%%%%%%%%%%%%%%%%%%%%%%%%%%
\section{Hand-crafted variable transformations for the one-site Hubbard model}
\label{sec:variable_transformations_hubbard}

PIK approaches are based on variable transformations: for a given distribution path $p_t$, infinitesimal (and subsequently global) variable transformations are directly computed via \labelcref{eq:phiMap,eq:PIKflow}. For well-established models such as the Hubbard model, however, hand-crafted transformations are known. The impact of exemplary hand-crafted variable transformations (see e.g.~\cite{Tanizaki:2015rda}) on the sign problem of the one-site Hubbard model is shown in \Cref{fig:average_sign_hubbard}. In this Appendix, we provide the explicit form of these variable transformations.

\textit{Original:} For completeness, we first note that, using simple Gau\ss ian identities, the distribution \labelcref{eq:HubbardZ} of the original one-site Hubbard model in the charge channel reads
\begin{align}
	p(\varphi) = &\,\frac{1}{\mathcal{N}} \Big[ \, e^{-\frac{\beta}{2U} \varphi^2} + 2 e^{\beta \mu} \, e^{-\frac{\beta}{2U} (\varphi - iU)^2} \nonumber\\[1ex]
	&\qquad + e^{\beta(2\mu - U)} e^{-\frac{\beta}{2U} (\varphi - 2iU)^2} \,\Big]\,.
	\label{eq:distribution_hand-crafted_original}
\end{align}
As shown in \Cref{fig:average_sign_hubbard}, this formulation allows for real sampling if $\mu \lesssim -U/2$. There, the weight of the first Gau\ss ian, which corresponds to the empty state and is centred on the real axis, dominates.

\textit{Half filling:} At half filling ($\mu = U/2$) the transformation $\varphi' = \varphi - i U$ turns \labelcref{eq:distribution_hand-crafted_original} into
\begin{align}
	p(\varphi') &= \frac{1}{\mathcal{N}} \Big[ \, e^{-\frac{\beta}{2U} (\varphi' + iU)^2} \nonumber \\[1ex]
	& \qquad + 2 e^{\frac{\beta U}{2}} \, e^{-\frac{\beta}{2U} \varphi'^2} + e^{-\frac{\beta}{2U} (\varphi' - iU)^2} \,\Big] \nonumber \\[1ex]
	&= \frac{2\, e^{\frac{\beta U}{2}}}{\mathcal{N}} \, e^{-\frac{\beta}{2U} \varphi'^2} \Big[\,  1 + \cos(\beta\varphi')\,\Big]\,.
	\label{eq:distribution_hand-crafted_half-filling}
\end{align}
Here, we have combined the first and the last Gau\ss ian in the first line into the cosine in the second line. The distribution is now non-negative and can thus be interpreted as a probability distribution. At half filling, the average sign from \Cref{fig:average_sign_hubbard} quantifies the quality of reweighting from the fixed $\mu=U/2$ to the other chemical potential values shown on the x-axis.

\textit{Particle-hole:} For the particle-hole transformation, one may use the transformation $\varphi' = \varphi - 2iU$. Then, the Gau\ss ian corresponding to the doubly occupied state yields real weights and one obtains
\begin{align}
	p(\varphi') = &\,\frac{1}{\mathcal{N}} \Big[ \, e^{-\frac{\beta}{2U} (\varphi' + 2iU)^2} + 2 e^{\beta \mu} \, e^{-\frac{\beta}{2U} (\varphi' + iU)^2}  \nonumber\\*[1ex]
	&\qquad + e^{\beta(2\mu - U)} e^{-\frac{\beta}{2U} \varphi'^2} \,\Big]\,.
\end{align}
We conclude this Appendix by noting that this hand-crafted variable transformation is indeed found blindly by the PIKfold in \Cref{fig:pik_particle_hole_contour}, where it is applicable.

%%%%%%%%%%%%%%%%%%%%%%%%%%%%%
\section{The fate of a distribution's zeros under the PIKfan transport}
\label{sec:zeros_and_transport}
In this paper and its Appendices, we have illustrated with concrete examples how zeros of the evolving distributions $p_t$ restrict measure transport approaches along a single contour deformation, recall \Cref{sec:RippingExample,sec:OneSiteHubbard}. This restriction is resolved constructively by the PIKfan, see e.g.~\Cref{sec:PIKfanExample,sec:phi4_mixture}. In particular, under the transport via the PIKfan, zeros of the distribution $p_t$ may lie on the sampling manifolds $\mathcal{M}_{i,t}$ without necessarily causing the ripping encountered for a single contour deformation. An example of this is the transport of the one-site Hubbard model in \Cref{fig:hubbard_mixture_manifolds_3d}.

In this Appendix, we explain the underlying mechanism. Once again, the central property that resolves the issues with the distribution's zeros is the weight preservation of the PIK, and the central elements of this description are already provided in~\cite{Ihssen:2026njd}. The weight-preserving property of PIKs tells us that $d\mu_t\, p_t(\phi_t)$ is constant during the flow. In particular, a zero $p_t(\phi_t) = 0$ must remain a zero of the distribution during the flow. Similarly, for a regular $d\mu_t$, values of the distribution $p_t$ that are non-zero must retain non-zero distribution values for a regular flow. In the following, we explain this property in terms of differential equations for $p_t$. Thus, the following discussion can be seen as an exercise in the application of the weight-preserving property.

As described in~\cite{Ihssen:2026njd}, the weight-preserving transport of a smooth holomorphic distribution $p_t: \mathbb{C}^\mathcal{D} \to \mathbb{C}$ is provided by the Wegner equation
\begin{align}
	\frac{ \partial p_t(\phi)}{\partial t} = - \frac{\partial}{\partial \phi_{j}} \left[ \dot\phi_{t,j}(\phi) \, p_t(\phi) \right]\,,
	\label{eq:wegner_zeros_transport_app}
\end{align}
where the fields are transported by the kernel vector field $\dot\phi_t: \mathbb{C}^\mathcal{D} \to \mathbb{C}^\mathcal{D}$ via
\begin{align}
	\frac{d \phi_t}{dt} = \dot\phi_t(\phi_t) \,.
	\label{eq:field_transport_zero_transport_app}
\end{align}
Using \labelcref{eq:wegner_zeros_transport_app} and basic properties of the total derivative, the distribution along the field transport with \labelcref{eq:field_transport_zero_transport_app} is then provided by
\begin{align}
	\frac{d p_t(\phi_t)}{dt} &= \frac{\partial p_t(\phi_t)}{\partial t} + \dot\phi_{t,j}(\phi_{t}) \frac{\partial p_t(\phi_t)}{\partial\phi_j} \nonumber \\[1ex]
	&\overset{\labelcref{eq:wegner_zeros_transport_app}}{=} - \frac{\partial \dot\phi_{t,j}(\phi_t)}{\partial \phi_j} p_t(\phi_t) \,.
	\label{eq:dpdt_along_transport}
\end{align}
Solving \labelcref{eq:dpdt_along_transport} in time along the transport of the fields thus yields
\begin{align}
	p_t(\phi_t) = p_0(\phi_0) \exp\left[- \int_0^t ds\, \frac{\partial \dot\phi_{s,j}(\phi_s)}{\partial \phi_j} \right] \,,
	\label{eq:pt_along_transport}
\end{align}
which follows directly from the weight-preserving property of the PIK and connects it to the change-of-variables formula for the distribution.

\Cref{eq:pt_along_transport} connects the evolution of the distribution's zeros with the regularity of the kernel $\dot\phi_t$. For a regular flow, the integral in the exponent is finite. Hence, in finite time, the flow can neither map a non-zero value of $p_t$ onto zero, nor a zero onto a non-zero value. In other words, a regular flow can neither create nor destroy zeros of $p_t$, and thus
\begin{align}
	\frac{d p_t(z_t)}{dt} = 0\,,
	\label{eq:zeros_preserved_along_transport}
\end{align}
where $z_t$ is a zero of the distribution $p_t$ that is transported along the flow via \labelcref{eq:field_transport_zero_transport_app}. 

Consequently, distribution paths that do not require zeros to be created or destroyed during the transport can result in perfectly regular kernels $\dot\phi_t$ that do not rip the manifold. In contrast, in the examples of \Cref{fig:FirstRIP,fig:individual_zero_rip}, the initial distribution has no zeros, whereas the later ones do. Hence, the integral in the exponent of \labelcref{eq:pt_along_transport} must diverge in finite time, and the kernel $\dot\phi_t$ becomes singular when $\mathcal{M}_t$ is pushed towards the zeros. Note that this reasoning based on the weight-preserving property is general and does not require the exact solution to the kernel. 

This reasoning holds in any dimension, while the geometry of the zeros changes with the number of lattice sites, $|\mathcal{D}|$. For holomorphic $p_t$, the zero set ${\cal Z}_t$ is a complex hypersurface. For $|\mathcal{D}|=1$ it consists of isolated points, which the sampling manifold generically meets only at isolated times $t$. For $|\mathcal{D}|\geq2$, in contrast, ${\cal Z}_t$ is extended, and a transversal intersection with ${\cal M}_t$ is $(|\mathcal{D}|-2)$-dimensional and stable under small deformations of the manifold. Hence, for $|\mathcal{D}|\geq2$, zeros generically lie on the sampling manifold over finite intervals of the RG-time. This causes no ripping as long as the flow does not have to create them, which is ensured by an initial distribution $p_0$ that already carries the relevant zeros, as provided by the PIKfan.

\Cref{eq:pt_along_transport} also shows that the exponent is a natural scalar probe for the ripping of the manifold. Its modulus $\Lambda_t(\phi_0)$ from \labelcref{eq:regularity_indicatorApp} is already used in \Cref{sec:NoRip,sec:PIKfanCauchy}. The real part of the exponent is $\ln|p_t(\phi_t)/p_0(\phi_0)|$, which diverges when a sample approaches a zero of $p_t$, while its imaginary part records the change of the phase of the distribution along the flow.

For a distribution path that neither creates nor destroys zeros, \labelcref{eq:zeros_preserved_along_transport} allows us to directly compute the flow of zeros,
\begin{align}
	0 &= \frac{d p_t(z_t)}{dt} = \frac{\partial p_t(z_t)}{\partial t} + \dot\phi_{t,j}(z_t) \frac{\partial p_t(z_t)}{\partial \phi_{j}} \,.
	\label{eq:zeros_along_transport}
\end{align}
This is not a differential but merely a linear equation for the kernel at the zeros. Thus, \labelcref{eq:zeros_along_transport} directly allows one to compute the flow of the zeros along the path $p_t$ under the requirement that $\dot\phi_t$ should stay regular and holomorphic.

Finally, as described in \Cref{sec:RealSampling}, the PIKfan allows real sampling where a distribution path with a single contour deformation fails. A crucial step in this construction is that the same kernel $\dot\phi_t^{(w)}$, computed for the full distribution path $p_t$, transports all distributions $\rho_{i,t}$ of the PIKfan. Applying \labelcref{eq:pt_along_transport}  to the split distributions $\rho_{i,t}$ gives
\begin{align}
	\rho_{i,t}(\phi_t) &= \rho_{i,0}(\phi_0) \exp\left[- \int_0^t ds\, \frac{\partial \dot\phi_{s,j}^{(w)}(\phi_s)}{\partial \phi_j} \right] \,, 
\end{align}
where $i$ labels the distributions and $j$ labels the lattice points. 
Thus, for a regular flow, one obtains
\begin{align}
	\rho_{i,t}(\phi_t) = \rho_{i,0}(\phi_0) \frac{p_t(\phi_t)}{p_0(\phi_0)} \,.
	\label{eq:relation_ribs_and_full_distribution}
\end{align}
Here, the ratio in \labelcref{eq:relation_ribs_and_full_distribution} shows yet again that encountering a zero in $p_t$ may still lead to perfectly regular rib distributions $\rho_{i,t}$. This is the case if zeros of $p_0$ are compensated by zeros of $p_t$ for all $t$ and the flow map $\phi_t$ remains injective. The latter is required to avoid branch points, see \Cref{sec:PIKfanCauchy}.

%%%%%%%%%%%%%%%%%%%%%%%%%%%%%%%%%%%%%%

	\bibliographystyle{apsrev4-2}
	\bibliography{refs}
    
\end{document}